%% file: main.tex
\documentclass[10pt,conference]{IEEEtran}
\IEEEoverridecommandlockouts

\usepackage{cite}
\usepackage{amsmath,amssymb,amsfonts}
\usepackage{algorithmic}
\usepackage{graphicx}
\usepackage{textcomp}
\usepackage[table]{xcolor}
\usepackage{subcaption}
\usepackage{multicol,multirow}
\usepackage{pifont}
\usepackage{enumitem}
\usepackage{makecell}
\usepackage{listings}
\usepackage{amsthm,thmtools}
\usepackage{soul}
\usepackage{xspace,fontawesome}
\usepackage{booktabs}
\usepackage{tcolorbox}

\setul{0.3ex}{0.2ex}

\def\BibTeX{{\rm B\kern-.05em{\sc i\kern-.025em b}\kern-.08em
    T\kern-.1667em\lower.7ex\hbox{E}\kern-.125emX}}

\newcommand{\MT}{{MetaReJ}}
\newcommand{\textcode}[1]{\texttt{#1}}

\definecolor{promptbg}{HTML}{F5F5F5}
\lstdefinestyle{lstprompt}{
    backgroundcolor=\color{promptbg},    
    basicstyle=\ttfamily\footnotesize,          
    commentstyle=\color{Green},      
    extendedchars=true,                  
    frame=single,                        
    keepspaces=true,                     
    keywordstyle=\color{blue},           
    breakatwhitespace=false,         
    breaklines=true,                 
    captionpos=b,                    
    keepspaces=true,                 
    showspaces=false,                
    showstringspaces=false,
    showtabs=false,                  
    tabsize=2,
    xleftmargin=3pt,
    xrightmargin=3pt,
    breakindent=10pt,
    postbreak=\mbox{\textcolor{gray}{$\hookrightarrow$}\space},
}

\definecolor{pblue}{rgb}{0.13,0.13,1}
\definecolor{pgreen}{rgb}{0,0.5,0}
\definecolor{pred}{rgb}{0.9,0,0}
\definecolor{pgrey}{rgb}{0.46,0.45,0.48}

\usepackage{listings}
\definecolor{promptgray}{rgb}{0.95,0.95,0.95}
\definecolor{variableorange}{rgb}{1.0,0.5,0}

\lstdefinelanguage{LLMPrompt}{
    basicstyle=\small\ttfamily,
    breaklines=true,
    numbers=none,
    frame=single,
    backgroundcolor=\color{promptgray},
    morekeywords={System, User, Assistant}, 
    keywordstyle=\bfseries\color{blue},
    morecomment=[l]{//}, 
    commentstyle=\color{pgreen}
}

\declaretheoremstyle[
  spaceabove=\topsep,
  spacebelow=-1\topsep,
  bodyfont=\normalfont\itshape,
  headformat=\NAME\NUMBER  
]{nospacetheorem}
\declaretheorem[style=nospacetheorem,name=MR]{mr}

\NewTColorBox{custombox}{o}{
    colback=gray!8,
	colframe=gray!25,
    boxrule=1pt,
	left=1.0mm,
	right=1.0mm,
	top=1.mm,
	bottom=1.mm,
	fonttitle=\bfseries,
	arc=0mm,
	notitle,
	before=\par\medskip\noindent,
    IfValueTF={#1}{before upper={\textbf{#1:}\space}}{},
    parbox=false,
}

\begin{document}

\title{Detecting Incorrect LLM Explanations for Automated Program Repair}

\makeatletter 
\newcommand{\linebreakand}{%
  \end{@IEEEauthorhalign}
  \hfill\mbox{}\par
  \mbox{}\hfill\begin{@IEEEauthorhalign}
}
\makeatother 

\author{\IEEEauthorblockN{Lam Nguyen Tung}
\IEEEauthorblockA{\textit{Monash University} \\
Melbourne, Australia \\
lam.nguyentung@monash.edu}
\and
\IEEEauthorblockN{Chenxi Gu}
\IEEEauthorblockA{\textit{Monash University} \\
Melbourne, Australia \\
chenxi.gu@monash.edu}
\and
\IEEEauthorblockN{Xiaoning Du}
\IEEEauthorblockA{\textit{Monash University} \\
Melbourne, Australia \\
xiaoning.du@monash.edu}
\and

\linebreakand 

\IEEEauthorblockN{Haoye Tian}
\IEEEauthorblockA{\textit{Aalto University} \\
Espoo, Finland \\
haoye.tian@aalto.fi}
\and
\IEEEauthorblockN{Aldeida Aleti}
\IEEEauthorblockA{\textit{Monash University} \\
Melbourne, Australia \\
aldeida.aleti@monash.edu}
\and
\IEEEauthorblockN{Neelofar Neelofar}
\IEEEauthorblockA{\textit{RMIT University} \\
Melbourne, Australia \\
neelofar.neelofar@rmit.edu}
}

\maketitle


\begin{abstract}
\input{sections/abstract}

\end{abstract}


\section{Introduction}
\label{sec:intro}
\input{sections/intro}

\section{Preliminary and Background}
\label{sec:preliminary}
\input{sections/preliminary}

\section{\MT: The Proposed Detection Method}
\label{sec:method}
\input{sections/method}

\section{Experimental Design}
\label{sec:evaluation}
\input{sections/RQs}
\input{sections/evaluation}

\section{Experimental Results}
\label{sec:results}

\input{sections/results}


\section{Threats to Validity}
\label{sec:threat2validity}
\input{sections/threats}

\section{Related Work}
\label{sec:related_work}
\input{sections/related_work}

\section{Conclusion}
\label{sec:Conclusion}
\input{sections/conclusion}

\bibliographystyle{IEEEtran}
\bibliography{reference}

\clearpage
%
\appendices
\input{sections/supplementary_material}

\end{document}

%% file: sections/abstract.tex
LLM-generated explanations 
are 
important in automated program repair (APR), as they enable developers to understand repair rationale and assess patch quality.
Their correctness can directly affect how developers respond to repairs.
For instance, 
incorrect explanations may signal 
unreliable repairs that fail to fix bugs.
Thus, detecting incorrect explanations is essential to help developers focus on reliable repairs. 
However, manual detection is impractical due to the growing number of LLM-generated repairs. 
Automated detection is therefore necessary to filter explanations presented to developers.

Prior methods mainly detect incorrect responses of general question-answering LLMs, not APR explanations.
They often sample responses and identify contradictions among them.
In APR, they face two limitations: (1)~to identify contradictions, they often rely on external verifiers, which may not exist or lack APR-specific reasoning; (2)~their sampling techniques, e.g., paraphrasing or same-input re-prompting, tend to produce superficial or 
redundant
samples. 
Such methods may reproduce the original incorrect explanation, 
failing to expose it via contradictions.

We propose \MT, a new metamorphic testing approach and the first detection method for incorrect APR explanations.
To avoid reliance on external verifiers, \MT{} adopts a metamorphic argumentative strategy inspired by debate and viva examination to challenge the LLM's explanation.
Using novel metamorphic relations (MRs), \MT{} recursively questions the LLM under equivalent and non-equivalent repair scenarios and captures contradictions in its responses.
To mitigate superficial sampling, \MT{} leverages program transformations.
It transforms the original buggy program into different variants that either preserve or alter the underlying bug.
It then samples explanations by asking the LLM how to fix the transformed variants. 
This technique produces more diverse and non-redundant samples,
making it easier to expose contradictions.
Such contradictions are encoded as MR violations.
If the LLM's responses violate the MRs,
\MT{} flags the original explanation as incorrect. 
To evaluate \MT{}, we constructed a manually-labeled dataset of 1,756 explanations of multiple LLM-based APR systems. 
Experiments show that \MT{} achieves up to 0.43 in MCC, outperforming prior methods by 333.3\% in MCC.

%% file: sections/intro.tex
Large language models (LLMs) have advanced automated program repair (APR), achieving state-of-the-art (SOTA) repair success rates of up to 70\% on existing benchmarks~\cite{Zhang:2026:RepairIngredientsAreAllYouNeed:ImprovingLargeLanguageModelBasedProgramRepairviaRepairIngredientsSearch, Xia:2024:AutomatedProgramRepairviaConversation:Fixing162outof337Bugsfor0.42EachusingChatGPT, Li:2025:EvaluatingtheGeneralizabilityofLLMsinAutomatedProgramRepair}.
Despite this, developers still prefer to remain involved in the repair process, mainly due to the lack of trust in fully automated repairs~\cite{Bohme:2017:WhereIstheBugandHowIsItFixedAnExperimentWithPractitioners, Winter:2023:HowdoDevelopersReallyFeelAboutBugFixingDirectionsforAutomaticProgramRepair}.
Building such trust requires developers to understand not only the generated patch, but also the rationale behind it~\cite{Meem:2024:ExploringExperienceswithAutomatedProgramRepairinPractice, Tao:2014:AutomaticallyGeneratedPatchesasDebuggingAids:AHumanStudy}.
LLM-provided explanations offer a useful mechanism for communicating the LLM's repair rationale to developers~\cite{Khati:2025:MappingtheTrustTerrain:LLMsinSoftwareEngineeringInsightsandPerspectives}.
This helps developers assess the quality of the generated repairs~\cite{Zhang:2024:VulAdvisor:NaturalLanguageSuggestionGenerationforSoftwareVulnerabilityRepair}.
The correctness of explanations may influence how developers act on the repairs.
For example, incorrect explanations
indicate unreliable repairs that fail to fix bugs and should be disregarded.
Hence, detecting incorrect explanations is essential in LLM-based APR.
However, it is overwhelming for developers to detect them manually, as the number of LLM-generated repairs continues to grow extensively~\cite{Zhang:2026:RepairIngredientsAreAllYouNeed:ImprovingLargeLanguageModelBasedProgramRepairviaRepairIngredientsSearch, Xia:2024:AutomatedProgramRepairviaConversation:Fixing162outof337Bugsfor0.42EachusingChatGPT}.
This creates a need for automated detection of incorrect explanations.
Such automation reduces the number of explanations presented to developers, alleviates review burden, and facilitates the repair process.

No available method detects incorrect APR explanations. 
Most methods identify if the final answers of LLMs in general question-answering tasks are incorrect. 
These methods often sample multiple responses and identify if these responses contradict one another~\cite{Manakul:2023:SelfCheckGPT:ZeroResourceBlackBoxHallucinationDetectionforGenerativeLargeLanguageModels, Yang:2025:HallucinationDetectioninLargeLanguageModelswithMetamorphicRelations}.
For example, SelfCheckGPT~\cite{Manakul:2023:SelfCheckGPT:ZeroResourceBlackBoxHallucinationDetectionforGenerativeLargeLanguageModels} re-prompts the LLM with the same query and compares the new responses with the original using semantic similarity metrics or external LLMs.
MetaQA~\cite{Yang:2025:HallucinationDetectioninLargeLanguageModelswithMetamorphicRelations} and DrHall~\cite{Wu:2025:DetectingandReducingtheFactualHallucinationsofLargeLanguageModelswithMetamorphicTesting} paraphrase the response and identify contradictions among the paraphrased versions.

Unfortunately, existing methods are not well-suited or effective for APR explanations.
We identify two key limitations.
\textbf{\textit{First}}, methods like SelfCheckGPT rely on external verifiers to identify contradictions.
In APR, such verifiers often do not exist or lack APR-specific knowledge.
\textbf{\textit{Second}}, existing methods sample explanations by re-prompting LLMs with the same input or by paraphrasing the explanation.
Such sampled explanations are often 
low-diversity or only superficially different at the word level.
They may preserve incorrectness rather than revealing it through contradictions~\cite{Yang:2025:HallucinationDetectioninLargeLanguageModelswithMetamorphicRelations}.

Exhaustively detecting incorrect explanations is a non-trivial task. 
Hence, we take a step toward mitigating manual review effort and facilitating the APR process.
We achieve this by mitigating the two limitations.
We propose \MT{}, a new metamorphic testing method and the first approach for detecting incorrect APR explanations.
To mitigate the \textbf{\textit{first}} limitation, we adopt a metamorphic argumentative strategy to challenge the LLM's explanation.
Rather than relying on external verifiers, \MT{} probes whether the LLM can defend itself under a series of recursive questions without providing contradictory responses.
Intuitively, reliable reasoning for generating a correct explanation should transfer to and support these probing questions.
In contrast, unreliable reasoning is more likely to surface as contradictions.

This argumentative strategy is inspired by debate and viva examination.
A given claim is not directly judged by outside knowledge or understanding.
The adopted strategy treats the claim as a \textit{premise}.
The claim-giver then responds to a series of recursive questions to determine whether later responses reinforce or contradict that premise. 
In this way, the claim-giver is challenged using their own claims, assumptions, or logic.
Analogously, 
we treat the LLM as the claim-giver. 
The original explanation and patch are considered the claim.
\MT{} questions the LLM to generate sampled explanations and then to verify if these explanations apply to the original program.
Metamorphic relations (MRs) are introduced to design such questions and represent the expected responses.

To mitigate the \textbf{\textit{second}} limitation, 
we do not apply prior superficial sampling techniques. Instead,
we apply diverse, non-redundant transformations at the program level.
We transform the original buggy program in terms of syntax or semantics.
We then sample explanations by asking the LLM how to fix the transformed programs.
These explanations, referred to as follow-up explanations, will inevitably be transformed accordingly.
This allows \MT{} to mitigate the repetition of superficial or redundant sampled explanations.

We encode program transformations in two MRs to generate and evaluate follow-up explanations.
[\textit{MR}$_1$]~Explanations for fixing semantically \textit{equivalent} programs 
\textit{apply} to the original program. 
[\textit{MR}$_2$]~Explanations for fixing semantically \textit{non-equivalent} programs should \textit{not apply}, as they have different bugs.
The LLM is expected to satisfy these MRs.

Given these two MRs, \MT{} performs the metamorphic argumentative strategy in three steps.
First, it transforms the original buggy program using semantic-preserving and semantic-altering transformations.
Second, it asks the LLM to generate follow-up explanations for fixing the transformed programs.
Third, it asks the same LLM to evaluate if these explanations satisfy the MRs.
Our key insight is that the LLM's explanation is incorrect if the LLM's responses are likely to violate the MRs.
Such violations mean the LLM contradicts itself about fixing the buggy program, likely to generate an incorrect explanation.

To effectively adopt the metamorphic argumentative strategy, we also explicitly provide the LLM with the original patch as the context when generating follow-up explanations.
This is the critical step: it prepares the LLM to be examined, sets up the \textit{premise}, and anchors follow-up explanations to the same original patch. 
By doing so, contradictions can be unambiguously traceable back to the original explanation. 
Otherwise, the strategy may freely generate follow-up explanations that introduce brand new claims.
These explanations are therefore less commensurable with the original one. 
This makes it difficult to determine if apparent contradictions are truly due to the original explanation or to the new claims.
Hence, this weakens the logical foundation of the metamorphic argumentative strategy, as used in debate or viva examination.

As detecting incorrect APR explanations is not trivial, \MT{} is designed to be high-precision while increasing the chances of detecting them.
Its metamorphic argumentative strategy can improve precision by avoiding external verifiers, which introduce a new source of uncertainty that may cause false contradictions.
\MT{} derives contradictions from the LLM's own reasoning, providing stronger evidence of incorrect explanations.
The improvement in precision, however, comes with a potential loss in recall.
\MT{} may fail when the LLM is stubborn and consistently defends its incorrect reasoning.
By using program transformations, we mitigate the second limitation but also compensate for the loss in recall and further strengthen it by diversifying the sampling.

To evaluate \MT{}, we create a manually-labeled dataset of APR explanations.
These explanations are collected from three SOTA systems: ThinkRepair~\cite{Yin:2024:ThinkRepair:SelfDirectedAutomatedProgramRepair}, ReInFix~\cite{Zhang:2026:RepairIngredientsAreAllYouNeed:ImprovingLargeLanguageModelBasedProgramRepairviaRepairIngredientsSearch}, and MORepair~\cite{Yang:2026:MORepair:TeachingLLMstoRepairCodeviaMultiObjectiveFineTuning} on the respective bug datasets used in the original studies.
These studies had already mitigated data leakage.
We formulated the detection of incorrect explanations as a binary classification problem.
Our experiments show that \MT{} achieves on average Precision of 0.92, Recall of 0.64, and MCC (Matthew Correlation Coefficient~\cite{Chicco:2023:TheMatthewsCorrelationCoefficientMCCShouldReplacetheROCAUCastheStandardMetricforAssessingBinaryClassification}) of 0.39.
It also outperforms the SOTA methods designed for general question-answering by up to 30.0\% in Precision, 536.4\% in Recall, and 333.3\% in MCC.

Our main contributions can be summarised as follows:
\begin{itemize}
    \item The first benchmark of the correctness of explanations generated by LLM-based APR systems,
    \item \MT, the first approach that detects incorrect explanations of LLM-based APR systems,
    \item A comprehensive evaluation of \MT{} on the constructed benchmark, which shows that \MT{} improves methods designed for general question-answering by up to 30.0\% in Precision, 536.4\% in Recall, 333.3\% in MCC, 
    \item The dataset and implementation are available at \textbf{https://doi.org/10.5281/zenodo.21059953}.
\end{itemize}

%% file: sections/preliminary.tex

We show an example to motivate \MT. 
We then introduce metamorphic testing, which forms a metamorphic argumentative strategy inspired by debate and viva examination.

\subsection{Motivating Example} \label{sec:Motivation}
\begin{figure*}[t]
    \centering
    \includegraphics[width=\linewidth]{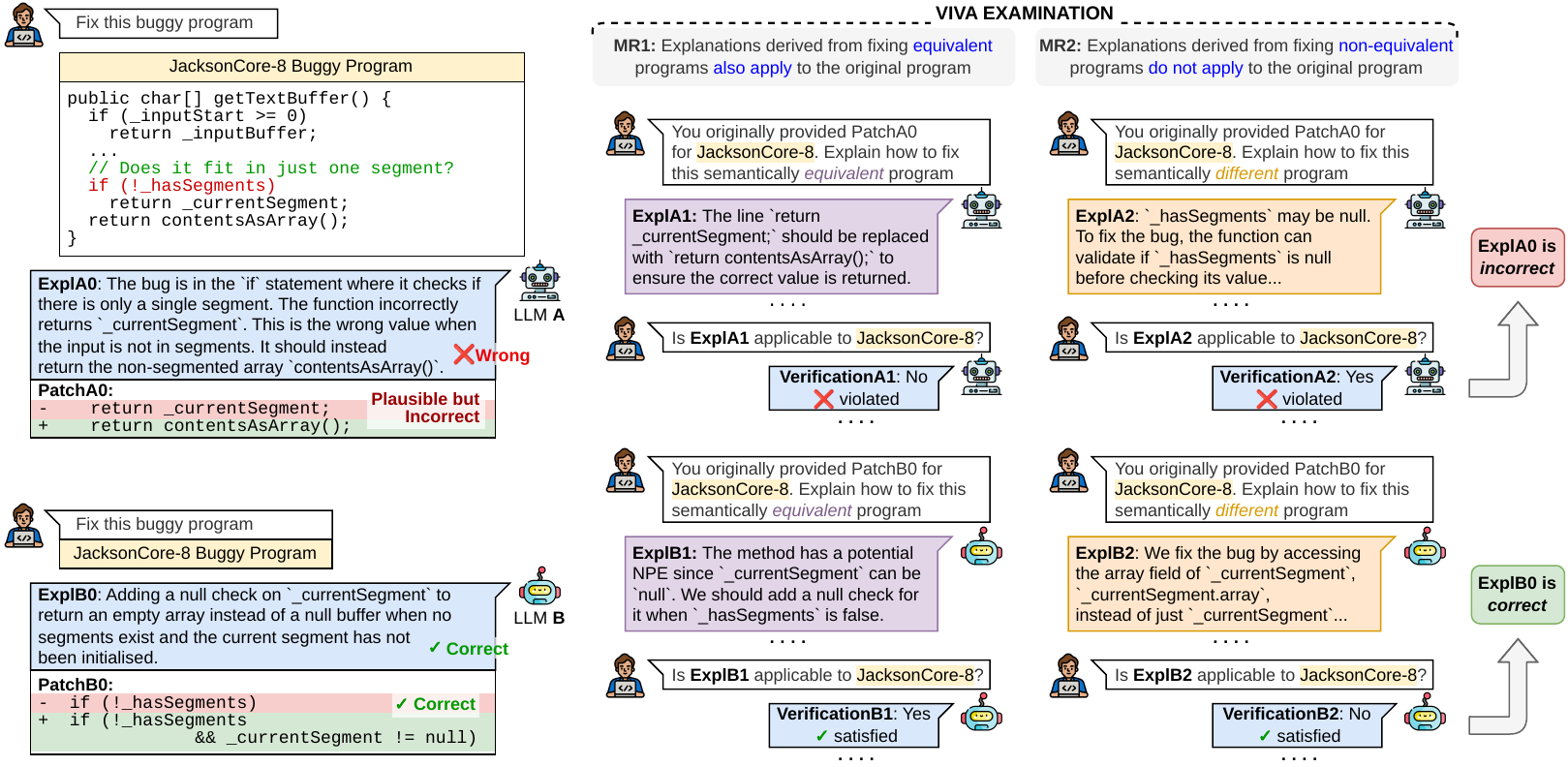}
    \caption{A motivating example: Two different repairs A and B (left side) generated by ThinkRepair~\cite{Yin:2024:ThinkRepair:SelfDirectedAutomatedProgramRepair} and ReInFix~\cite{Zhang:2026:RepairIngredientsAreAllYouNeed:ImprovingLargeLanguageModelBasedProgramRepairviaRepairIngredientsSearch}, respectively, for fixing JacksonCore-8 in Defects4J~\cite{Just:2014:Defects4j:ADatabaseofExistingFaultstoEnableControlledTestingStudiesforJavaPrograms}.
    Repair A provides an incorrect explanation ExplA0 and a plausible but incorrect patch.
    Repair B provides both a correct explanation ExplB0 and a correct patch.
    \MT{} can classify ExplA0 as incorrect and ExplB0 as correct (right side).}
    \label{fig:motivating_example}
\end{figure*}

The left side of Figure~\ref{fig:motivating_example} shows the buggy program \textcode{getTextBuffer}, also known as JacksonCore-8 in Defects4J~\cite{Just:2014:Defects4j:ADatabaseofExistingFaultstoEnableControlledTestingStudiesforJavaPrograms}.
The bug arises in how the single-segment case is handled \textcode{if} \textcode{(!\_hasSegments)} \textcode{return} \textcode{\_currentSegment;}.
When only one segment exists, the program returns \textcode{\_currentSegment} without checking if it is null.
To fix JacksonCore-8, ThinkRepair~\cite{Yin:2024:ThinkRepair:SelfDirectedAutomatedProgramRepair} and ReInFix~\cite{Zhang:2026:RepairIngredientsAreAllYouNeed:ImprovingLargeLanguageModelBasedProgramRepairviaRepairIngredientsSearch}, two SOTA LLM-based APR systems, generate two repairs A and B, respectively.
In each repair, the system generates a patch and an explanation.

As expressed in natural language, the explanations are useful to help developers understand the repair rationale and review patch quality.
ThinkRepair's explanation, ExplA0, reveals its flawed reasoning, suggesting that this repair is unreliable.
The \textcode{if (!\_hasSegments)} branch is supposed to handle the single-segment case.
ThinkRepair incorrectly thinks that returning \textcode{\_currentSegment} is buggy in this case.
In fact, if there is only one segment, \textcode{\_currentSegment} may truly contain the correct content, and returning it is correct.
On the other hand, ReInFix's explanation, ExplB0, is correct.
It correctly identifies the potential null pointer exception.
The suggestion in this explanation is also correct, adding a null check on \textcode{\_currentSegment} before returning it.

Based on the explanations, ReInFix's repair is clearly more reliable than ThinkRepair's.
Indeed, ThinkRepair's patch turns out to be incorrect, although it is plausible by passing all available test cases.
In this case, the test cases may be incomplete and the patch is overfitting.
In contrast, ReInFix's patch is correct, successfully eliminating the bug.

This example demonstrates that incorrect explanations of LLM-based APR systems help developers disregard unreliable repairs.
However, manually detecting them requires a lot of effort and is inherently impractical, given the growing number of LLM-generated repairs.
This motivates the need for automated detection methods like \MT{} to reduce developer effort.
\label{sec:Example}

The right side of Figure~\ref{fig:motivating_example} illustrates \MT's metamorphic argumentative strategy, where it correctly classifies ExplA0 as incorrect and ExplB0 as correct.
First, it applies semantic-preserving and semantic-altering transformations to \textcode{getTextBuffer}
to produce multiple transformed buggy programs Program1, Program2, ..., ProgramN\footnote{Due to page limitation, we illustrate only one equivalent transformed program and one non-equivalent transformed program}.
For instance, \MT{} generates semantically equivalent Program1 by rewriting the code using nested if–else blocks.
In contrast, \MT{} generates semantically non-equivalent Program2 by changing \textcode{\_inputStart} \textcode{>=} \textcode{0} to \textcode{\_inputStart} \textcode{>} \textcode{0}.
Details about these transformed programs can be found in Supplementary Material~\ref{sec:Appendix_Example}.

Under the two MRs, \MT{} asks each APR system to explain how to fix the transformed programs.
For example, ThinkRepair generates ExplA1 for Program1 and ExplA2 for Program2.
ReInFix generates ExplB1 for Program1 and ExplB2 for Program2.
ExplA1 and ExplB1 describe the same rationale as ExplA0 and ExplB0, respectively, while ExplA2 and ExplB2 illustrate rationales different from them.  

\MT{} then asks each system if each follow-up explanation applies to the original \textcode{getTextBuffer}.
We expect these systems to think that ExplA1 and ExplB1 also fix \textcode{getTextBuffer}, and ExplA2 and ExplB2 do not.
However, ThinkRepair fails to satisfy these relations when assessing ExplA1 and ExplA2.
It incorrectly answers ``No'' for ExplA1, thinking it is not applicable to \textcode{getTextBuffer}.
It also incorrectly answers ``Yes'' for ExplA2, even though ExplA2 is derived from fixing a different bug.
Since all ThinkRepair's responses violate the two MRs, its violation score is 1.
\MT{} classifies ExplA0 as incorrect.
In contrast, ReInFix's responses consistently satisfy the MRs.
ReInFix answers ``Yes'' for ExplB1 and ``No'' for ExplB2, as expected.
Hence, its violation score is 0 and \MT{} classifies ExplB0 as correct.

\subsection{Metamorphic Testing}
\label{sec:MetamorphicTesting}

Metamorphic testing~\cite{Chen:1998:MetamorphicTesting:ANewApproachforGeneratingNextTestCases} addresses the oracle problem~\cite{Barr:2015:TheOracleProbleminSoftwareTesting:ASurvey}, where expected outputs are unavailable.
Instead of directly verifying output correctness, it generates follow-up inputs and checks whether the resulting outputs satisfy predefined MRs. 
In this setting, the test oracle becomes if the MRs are satisfied.

Formally, an MR $R$ is a property between a set of inputs $X$$=$$\{ x_1, x_2, ..., x_n \}$ and their outputs $Y$$=$$\{$$f(x_1),$$f(x_2), ...,$ $f(x_n) \}$ for a function $f$.
It can be denoted as $R$$\subseteq$$X^n$$\times$$Y^n$~\cite{Barr:2015:TheOracleProbleminSoftwareTesting:ASurvey}.
Essentially, an MR specifies how a system should behave under specific transformations to the inputs. 
If the system's outputs violate the MRs,  
it indicates that the system may produce incorrect outputs.

MRs have been applied in traditional software testing~\cite{Barr:2015:TheOracleProbleminSoftwareTesting:ASurvey, Segura:2016:ASurveyonMetamorphicTesting}.
More recently, research has defined MRs to classify the factual correctness of information produced by NL question-answering systems, including LLMs~\cite{Wu:2025:DetectingandReducingtheFactualHallucinationsofLargeLanguageModelswithMetamorphicTesting, Yang:2025:HallucinationDetectioninLargeLanguageModelswithMetamorphicRelations, Cho:2025:MetamorphicTestingofLargeLanguageModelsforNaturalLanguageProcessing, Guo:2026:MORTAR:MultiTurnMetamorphicTestingforLLMBasedDialogueSystems}.
However, their application to explanation correctness, particularly in coding tasks like APR, remains underexplored.

As far as we know, this work is the first to leverage MRs to detect incorrect APR explanations.
Inspired by debate and viva examination, we introduce a metamorphic argumentative strategy that exposes contradictions in an LLM's responses.
MRs design recursive questions to challenge the LLM and encode its expected responses.
MR violations represent contradictions in the LLMs' bug-fixing reasoning.
Such violations are more likely when the original explanation is incorrect.

%% file: sections/method.tex

In this section, we describe \MT{}, the first method for detecting incorrect APR explanations.
\MT{} adopts a metamorphic argumentative strategy.
This strategy is inspired by debate and viva examination, where a claim acts as a premise under examination.
In APR, the LLM is considered the claim-giver; the original repair is treated as the claim.

Let $m$ represent the LLM-based APR system under test.
We denote by \textcode{Prog} the set of programs that implement a function, and by \textcode{Expl} the set of textual explanations that describe LLM reasoning.
Let $x \in \text{\textcode{Prog}}$ be the buggy program, $r \in \text{\textcode{Expl}}$ be the original explanation, and $y \in \text{\textcode{Prog}}$ be the original patch.
In this case, the repair $(r, y)$ can be considered as the given claim.
Given this repair, \MT's goal is to detect if the generated explanation $r$ is incorrect.

\begin{figure}
    \centering
    \includegraphics[width=\linewidth]{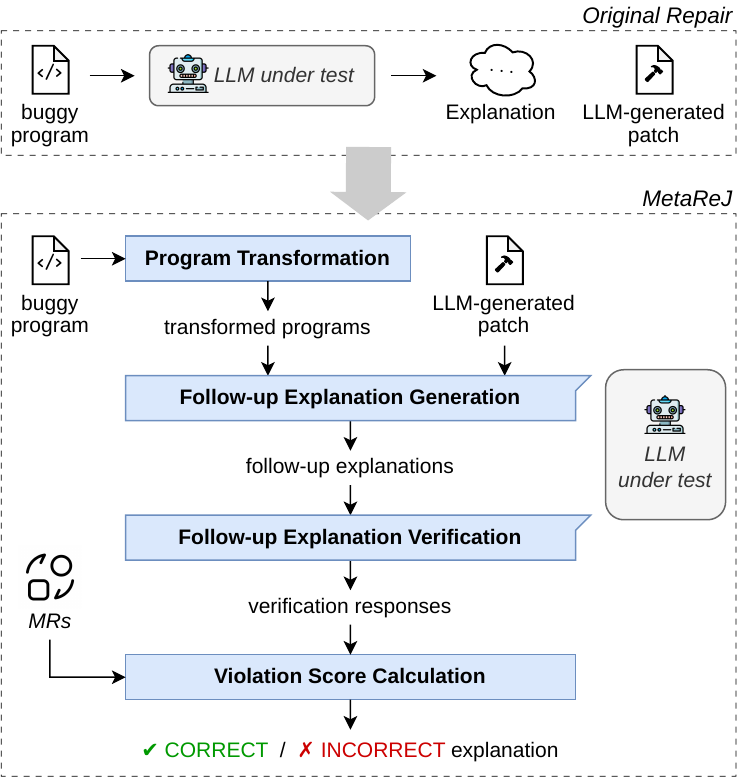}
    \caption{An explanation and patch are generated by the LLM under test for fixing the buggy program. 
    \MT{} performs the metamorphic argumentative strategy to determine if the explanation is incorrect.
    First, it transforms the buggy program into multiple equivalent and non-equivalent variants.
    Next, \MT{} asks the LLM to explain how to fix each transformed variant, given the originally generated patch as the argumentative context.
    This produces a set of follow-up explanations.
    \MT{} then asks the LLM whether each follow-up explanation applies to the original buggy program.
    Based on whether the LLM's responses violate the MRs, \MT{} calculates a violation score and uses it to detect if the original explanation is incorrect.
    }
    \label{fig:method}
\end{figure}
Figure~\ref{fig:method} illustrates the process of \MT.
It recursively questions the LLM under equivalent and non-equivalent repair scenarios, and exposes contradictions in the LLM's responses.
The process consists of four main steps: program transformation, follow-up explanation generation, follow-up explanation verification, and violation score calculation.
These steps are described in \ref{sec:ProgramTrans}, \ref{sec:FollowupRepair}, \ref{sec:FollowupVerify}, and \ref{sec:ScoreCalc}, respectively.

\subsection{Program Transformation}
\label{sec:ProgramTrans}

This step aims to transform the original buggy program $x$ into a diverse set of semantically equivalent and non-equivalent buggy programs.
We define $\text{\textcode{Tran}}_{\text{\textit{eqv}}}$ and $\text{\textcode{Tran}}_{\text{\textit{neq}}}$ for preserving and altering the semantics of a program, respectively.
They are modeled as two functions that map each program $x \in \text{\textcode{Prog}}$ to a corresponding set over \textcode{Prog}.
Formally, for all $x \in \text{\textcode{Prog}}$:
\begin{equation}
    \text{\textcode{Tran}}_{\text{\textit{eqv}}} : x \mapsto \{x'_{\text{\textit{eqv}}} \in \text{\textcode{Prog}} \mid x \equiv x'_{\text{\textit{eqv}}}\}
\end{equation}
\begin{equation}
    \text{\textcode{Tran}}_{\text{\textit{neq}}} : x \mapsto \{x'_{\text{\textit{neq}}} \in \text{\textcode{Prog}} \mid x \not\equiv x'_{\text{\textit{neq}}}\}
\end{equation}
A transformed program sampled from $\text{\textcode{Tran}}_{\text{\textit{eqv}}}$ or $\text{\textcode{Tran}}_{\text{\textit{neq}}}$ for $x$ is thus modeled by a variable as follows:
\vspace{-8mm}
\begin{multicols}{2}
\begin{equation}
x'_{\text{\textit{eqv}}} \in \text{\textcode{Tran}}_{\text{\textit{eqv}}}(x)
\end{equation}
~
\begin{equation}
x'_{\text{\textit{neq}}} \in \text{\textcode{Tran}}_{\text{\textit{neq}}}(x)
\end{equation}
\end{multicols}
\vspace{-2mm}
\noindent where $x'_{\text{\textit{eqv}}}$ is semantically equivalent to $x$, sharing the same underlying bug, while $x'_{\text{\textit{neq}}}$ is not.

Program transformations have been extensively studied in prior research, including APR~\cite{Thanh:2025:TowardsReliableEvaluationofNeuralProgramRepairwithNaturalRobustnessTesting}.
Hence, we reuse the existing transformations. 
Ultimately, we select 40 semantic-preserving and 20 semantic-altering transformations.
Supplementary Material~\ref{sec:AppendixTranSurvey} describes the detailed process of collecting transformations, and provides the list of transformations.



\MT{} \textit{aggregates} the selected transformations to generate transformed programs.
To validate the transformed programs, we compile and run them against the available test cases. 
Equivalent programs are expected to fail the same test cases as the original buggy program, whereas non-equivalent programs should fail different test cases. 
Programs that do not satisfy this criterion are discarded. 
The remaining programs are then used for follow-up explanation generation.

\subsection{Follow-up Explanation Generation}
\label{sec:FollowupRepair}

Given the curated transformed programs, \MT{} instructs the LLM $m$ to generate follow-up explanations. 
This process is designed to simulate the argumentative strategy used in debates or viva examinations.
To effectively adopt this strategy, the claim-giver must be aware of the claim as a premise.
Otherwise, 
the claim-giver may bring up new claims
that  
bear no logical relationship to the premise.
It is difficult to determine if contradictions stem from the original or the new claims.
This weakens the logical foundation of the argumentative strategy.
Hence, we do not repeat the original repair pipeline to produce follow-up explanations.
We design a repair sub-task that prepares the LLM for the argumentative strategy.
It explicitly provides the original patch $y$ as the context to set up the premise and anchor follow-up explanations to $y$. 

We denote this sub-task by $\text{\textcode{Explain}}_m(x', y)$ as a function that maps each pair of a transformed buggy program $x' \in \text{\textcode{Prog}}$ and the originally generated patch $y \in \text{\textcode{Prog}}$ to a follow-up explanation $r' \in \text{\textcode{Expl}}$. 
Formally, for all $(x', y) \in \text{\textcode{Prog}} \times \text{\textcode{Prog}}$:
\begin{equation}
    \text{\textcode{Explain}}_m : (x', y) \mapsto r' \in \text{\textcode{Expl}}
\end{equation}
where $x'=x'_{\text{\textit{eqv}}} \in \text{\textcode{Tran}}_{\text{\textit{eqv}}}(x)$ or $x'=x'_{\text{\textit{neq}}} \in \text{\textcode{Tran}}_{\text{\textit{neq}}}(x)$.

To implement $\text{\textcode{Explain}}_m(x', y)$, we embed the original patch $y$ directly into the prompt and ask $m$ to explain how to fix the transformed program $x'$ based on the given patch $y$. 
Supplementary Material~\ref{sec:PromptTemplate} shows the prompt template for follow-up explanation generation for transformed programs.

\subsection{Follow-up Explanation Verification}
\label{sec:FollowupVerify}

In this step, \MT{} asks the LLM under test $m$ to verify whether each follow-up explanation $r'$ applies to the original buggy program $x$.
We model this verification task performed by $m$ as $\text{\textcode{Verify}}_m$.
It is a function that maps a pair of the follow-up explanation $r' \in \text{\textcode{Expl}}$ and the buggy program $x \in \text{\textcode{Prog}}$ to 
$\mathbb{B} = \{0,1\}$. 
The value of 1 indicates that the model $m$ thinks the explanation $r'$ can be applied to the program $x$, and 0 otherwise.
Formally, for all $(r', x) \in \text{\textcode{Expl}} \times \text{\textcode{Prog}}$:
\begin{equation}
    \text{\textcode{Verify}}_m : (r', x) \in \text{\textcode{Expl}} \times \text{\textcode{Prog}}
    \mapsto 
    \{0,1\}
\end{equation}

This verification task can be performed through a straightforward question-answering process.
Specifically, \MT{} uses a prompt-based method that allows the LLM $m$ itself to rigorously assess whether the explanation is applicable to the buggy program.
The prompt template shown in Supplementary Material~\ref{sec:PromptTemplate} is used to guide the LLM in generating precise responses to enhance the overall reliability of \MT. 


If the original explanation $r$ is correct, the follow-up explanations derived from equivalent programs are expected to return ``Yes'' when verified, and those derived from non-equivalent programs should return ``No.'' 
When the LLM is uncertain about the correctness of a follow-up explanation, \MT{} instructs it to return ``Not Sure.''

\subsection{Violation Score Calculation}
\label{sec:ScoreCalc}

The goal of this step is to enhance the testing of LLM explanations by analyzing the violations of the two MRs. 
The MRs can be modeled as follows.
\begin{mr}
Explanations derived from equivalent programs that lead to the same patch also apply to the original program.
\end{mr}
\begin{equation}
\label{eq:MR1}
\text{\textcode{Verify}}_m(r'_{\text{\textit{eqv}}}, x) = 1
\end{equation}
where $r'_{\text{\textit{eqv}}} = \text{\textcode{Explain}}_m(x'_{\text{\textit{eqv}}}, y)$ and $x'_{\text{\textit{eqv}}} \in \text{\textcode{Tran}}_{\text{\textit{eqv}}}(x)$.
Violating MR1 means while $m$ fixes a program semantically equivalent to $x$ using $r'_{\text{\textit{eqv}}}$, $m$ does not think $r'_{\text{\textit{eqv}}}$ is applicable to $x$.

\begin{mr}
Explanations derived from non-equivalent programs do not apply to the original program.
\end{mr}
\begin{equation}
\label{eq:MR2}
\text{\textcode{Verify}}_m(r'_{\text{\textit{neq}}}, x) = 0
\end{equation}
where $r'_{\text{\textit{neq}}} = \text{\textcode{Explain}}_m(x'_{\text{\textit{neq}}})$ and $x'_{\text{\textit{neq}}} \in \text{\textcode{Tran}}_{\text{\textit{neq}}}(x)$.
Violating MR2 means that $m$ believes both $r'_{\text{\textit{neq}}}$ and $r$ are applicable to $x$, even though they are derived from fixing semantically different programs.

For the same LLM $m$, verifying explanations is inherently no more difficult than generating a correct one. 
This is because, from a complexity perspective~\cite{Papadimitriou:2003:ComputationalComplexity}, explanation generation requires $m$ to search over an extremely large space to produce a correct explanation. 
Therefore, if $m$ generates a correct explanation, it will 
provide verification results satisfying the expected outcome specified by the MRs.
Our empirical results in Supplementary Material~\ref{sec:AppendixProof} support this assumption.
When the LLM generates a correct explanation, it verifies explanations correctly with a probability of 0.90 (close to 1).
Moreover, verification failures are rare when the LLM generates correct explanations, while verification still often succeeds even when the LLM generates incorrect explanations.
These results show that verifying explanations is almost always no more difficult than generating them.  
Hence, verification failures, which are encoded as MR violations, strongly imply that the LLM generates an incorrect explanation.
\begin{equation}
\begin{split}
\Big( &\text{\textcode{Verify}}_m(r'_{\text{\textit{eqv}}}, x) = 0 \; \lor \\
& \text{\textcode{Verify}}_m(r'_{\text{\textit{neq}}}, x) = 1 \Big) \implies \text{``} r \text{\textit{ is incorrect}''}
\end{split}
\end{equation}
We define the \textit{violation score} of the two MRs as follows.
\begin{align}
\begin{split}
\label{eq:Final}
\mathcal{V}_m(r, x) := \mathbb{P}\bigl( & \text{\textcode{Verify}}_m(r'_{\text{\textit{eqv}}}, x) = 0 
\\ & \lor \; \text{\textcode{Verify}}_m(r'_{\text{\textit{neq}}}, x) = 1 \bigl)
\end{split}
\end{align}
Based on this, the violation score provides a lower bound on the incorrectness probability:
\begin{equation} \label{eq:LowerBound}
\mathcal{V}_m(r, x) \leq \mathbb{P}(\text{``} r \text{\textit{ is incorrect}''} )
\end{equation}
Supplementary Material~\ref{sec:AppendixProof} provides detailed theoretical and empirical proofs.
In this study, we consider $r$ as incorrect if $\mathcal{V}_m(r, x) > \theta$, implying $\mathbb{P}(\text{``} r \text{\textit{ is incorrect}''} ) \geq \theta$. 
Consequently, this approach provides a statistically sound way to identify incorrect explanations without access to ground truths.

\subsubsection*{Practical consideration}
Program transformations $\text{\textcode{Tran}}_{\text{\textit{eqv}}}$ and $\text{\textcode{Tran}}_{\text{\textit{neq}}}$ are typically inexpensive.
However, generating follow-up explanations $\text{\textcode{Explain}}_m$ may be costly, as it requires querying the LLM multiple times. 
To reduce this cost, we adopt a fixed sampling budget strategy.
Given a budget $N \times 2$, we draw $N$ follow-up explanations for $N$ equivalent programs and $N$ follow-up explanations for $N$ non-equivalent ones.
To enhance the diversity of this sampling, \MT{} selects the most representative transformed programs.
It clusters each list of equivalent and non-equivalent transformed programs into $N$ clusters, and randomly chooses one in each cluster as a representative example.
Details about the clustering process are provided in Supplementary Material~\ref{sec:sample_programs}.
The value of $N$ will be analysed through a sensitivity analysis in RQ4.

Given the budget $N$, we define an \textit{empirical violation score} $\hat{\mathcal{V}}_m$ that approximates $\mathcal{V}_m$ described in Equation~\ref{eq:Final}.
\begin{align}
\begin{split}
\hat{\mathcal{V}}_m(r, x, N) = & \frac{1}{N \times 2} \Biggl( \sum^N_i \mathbb{I}\big[( \text{\textcode{Verify}}_m(r'_{\text{\textit{neq}}_i}, x)=1\bigl] 
\\ + & \sum^N_j \mathbb{I}\bigl[\text{\textcode{Verify}}_m(r'_{\text{\textit{eqv}}_j}, x) = 0 \bigl] \Biggl)
\end{split}
\raisetag{2em}
\end{align}
where 
$\mathbb{I}[\cdot]$ is the indicator function, which returns 1 when the predicate holds and 0 otherwise.
Let $v \in $ $\{\text{``Yes''},$ $\text{``Not Sure''},$ $\text{``No''}\}$ represent the verification response for the follow-up explanation $r'$. 
$\mathbb{I}[\cdot]$ only returns 0 or 1, while $v$ can receive three values.
To map these values, we design a mechanism that assigns a numeric score for each follow-up explanation on a scale of [0, 1] based on its verification response $v$. 
``Not Sure'' is considered equally likely to indicate whether the explanation is correct or not, thus receiving a score of 0.5. 
The scores are mapped as follows:
\begin{equation}
\mathbb{I}[\cdot] = \nonumber
\begin{cases}
1.0 \;\, \text{if } (r'_{\text{\textit{eqv}}} \text{ and } v = \text{No}) \text{ or } (r'_{\text{\textit{neq}}} \text{ and } v = \text{Yes}) \\
0.5 \;\, \text{if } v = \text{Not Sure} \\
0.0 \;\, \text{if } (r'_{\text{\textit{eqv}}} \text{ and } v = \text{Yes}) \text{ or } (r'_{\text{\textit{neq}}} \text{ and } v = \text{No}) \\
\end{cases}
\end{equation}

The empirical violation score approximates the original violation score while avoiding repeated expensive LLM queries. 
As $N$ increases, the estimation of the violation score becomes more faithful.
While this approximation introduces some additional variance, it is highly effective in practice~\cite{Manakul:2023:SelfCheckGPT:ZeroResourceBlackBoxHallucinationDetectionforGenerativeLargeLanguageModels, Yang:2025:HallucinationDetectioninLargeLanguageModelswithMetamorphicRelations}. 
It amortizes LLM sampling costs across many evaluations, enabling scalable violation estimation. 

According to Equation~\ref{eq:LowerBound}, we consider the original explanation $r$ as incorrect if $\hat{\mathcal{V}}_m > \theta$. 
The value of the threshold $\theta$ will be identified through sensitivity analysis in RQ4.

%% file: sections/RQs.tex


We study the following research questions (RQs):

\begin{enumerate}[label=\textbf{RQ\arabic*.}, leftmargin=27.5pt]
    \item To what extent do SOTA LLM-based APR systems produce incorrect explanations?
    \item How \textit{effective} is \MT{} in detecting incorrect explanations of LLM-based APR systems?
    \item How \textit{efficient} is \MT{} in detecting incorrect explanations of LLM-based APR systems?
    \item How do the parameters $N$ and $\theta$ impact \MT?
    \item How does each component influence \MT{}?
\end{enumerate}

To measure detection quality, we use standard binary classification metrics.
In particular, we adopt the Matthews Correlation Coefficient (\textit{MCC}) as it considers all four components of the confusion matrix, and produces more reliable statistical results than F1-score~\cite{Chicco:2023:TheMatthewsCorrelationCoefficientMCCShouldReplacetheROCAUCastheStandardMetricforAssessingBinaryClassification}.
We also report Balanced Accuracy (\textit{BA}), Precision (\textit{Pre}), Recall (\textit{Rec}), also known as True Positive Rate, and Specificity (\textit{Spe}), also known as True Negative Rate.
To measure efficiency, we report the number of input tokens, output tokens, and total tokens.

%% file: sections/evaluation.tex
\subsection{Hyperparameters}

In \MT, we set the LLM's temperature to 0 when generating and verifying follow-up explanations to enhance stability.
RQ4 investigates the optimal values of $N$ and $\theta$, while other RQs use their default values, 5 and 0.5, respectively.
The implementation of program transformations is reused from Le et al.'s work~\cite{Thanh:2025:TowardsReliableEvaluationofNeuralProgramRepairwithNaturalRobustnessTesting}.
We performed three runs per experiment.

\subsection{Baselines for Detecting Incorrect Explanations}

Several detection approaches rely on databases, search engines~\cite{Guo:2022:ASurveyonAutomatedFactChecking}, or internal access to LLMs~\cite{Sriramanan:2024:LLMCheck:InvestigatingDetectionofHallucinationsinLargeLanguageModels}.
As they are often unavailable, these approaches are unsuitable as baselines.
Consequently, we compare \MT{} with the following methods.

\begin{itemize}
    \item SelfCheckGPT~\cite{Manakul:2023:SelfCheckGPT:ZeroResourceBlackBoxHallucinationDetectionforGenerativeLargeLanguageModels}: an SOTA task-agnostic, zero-resource approach for checking the correctness of an LLM's response. 
    It repeatedly queries the LLM to generate multiple samples and measures their consistency with the original response. 
    Following the prior setting~\cite{Yang:2025:HallucinationDetectioninLargeLanguageModelswithMetamorphicRelations}, we set the LLM's temperature to 0.5 when sampling, which yields the best performance.

    \item MetaQA~\cite{Yang:2025:HallucinationDetectioninLargeLanguageModelswithMetamorphicRelations}: a method for detecting incorrect facts generated by an LLM for general question-answering.
    It paraphrases the originally generated fact, then checks if the paraphrases remain valid.
    For APR, 
    we adapt MetaQA by using the LLM under test to paraphrase the original explanation.
    We use the same setting as the original study.

    \item DrHall~\cite{Wu:2025:DetectingandReducingtheFactualHallucinationsofLargeLanguageModelswithMetamorphicTesting}: a metamorphic testing method for detecting incorrect facts generated by an LLM for general question-answering.
    Since most of its MRs are not applicable to APR,
    we only consider AMR1, which asks the LLM a follow-up question ``\textit{[Original Answer] Is this statement not true?}''
    We use the same setting as the original study.

    \item SelfCheck w/ ProgTrans: an extension of SelfCheckGPT. 
    It samples responses using transformed programs instead of re-querying the LLM with the same input. 
    It uses the same set of transformed programs as \MT.
    
\end{itemize}

\subsection{Studied Models (LLMs Under Test)}

We aim to select LLM-based APR systems published within the past five years at top venues in SE, such as conferences (e.g., ICSE, FSE, ASE, ISSTA) and journals (e.g., TOSEM, TSE, EMSE).
We focus specifically on systems that can produce explanations. 
At the time of writing, only a limited number of such systems are available, as most existing systems only generate patches. 
Three systems are selected:
\begin{itemize}
    \item ThinkRepair~\cite{Yin:2024:ThinkRepair:SelfDirectedAutomatedProgramRepair}: A self-directed LLM-based APR approach. ThinkRepair first gathers CoTs as repair knowledge. It then repairs bugs via few-shot example selection, iterative LLM interaction, and testing feedback.
    We use CodeLlama 13B as the underlying model, the best-performing open-source LLM in the original study.

    \item ReinFix~\cite{Zhang:2026:RepairIngredientsAreAllYouNeed:ImprovingLargeLanguageModelBasedProgramRepairviaRepairIngredientsSearch}:  An agentic framework that enables LLMs to autonomously search for repair ingredients throughout both the reasoning and solution phases of bug fixing.
    ReinFix employs GPT-4o as the underlying LLM.

    \item MORepair~\cite{Yang:2026:MORepair:TeachingLLMstoRepairCodeviaMultiObjectiveFineTuning}: A novel perspective on LLM fine-tuning for APR. 
    It not only adapts model parameters to the syntactic nuances of code transformation but also fine-tunes the model to capture the underlying logic behind code changes in the training data.
    We use CodeLlama 13B as the underlying LLM, as in the original study.
\end{itemize}

\subsection{Benchmark Construction}
\label{sec:benchmark}

\input{sections/reasoning_benchmark}

%% file: sections/reasoning_benchmark.tex
To answer \textbf{RQ1}, we constructed a benchmark of the correctness of explanations generated by LLM-based APR systems. 

\subsubsection{Data leakage}
All three original studies~\cite{Yin:2024:ThinkRepair:SelfDirectedAutomatedProgramRepair, Zhang:2026:RepairIngredientsAreAllYouNeed:ImprovingLargeLanguageModelBasedProgramRepairviaRepairIngredientsSearch, Yang:2026:MORepair:TeachingLLMstoRepairCodeviaMultiObjectiveFineTuning} evaluated their APR systems on Defects4J~\cite{Just:2014:Defects4j:ADatabaseofExistingFaultstoEnableControlledTestingStudiesforJavaPrograms}.
To mitigate potential data leakage, they also evaluated their systems on benchmarks constructed after the LLMs' cutoff dates.
ThinkRepair and ReinFix were evaluated on RWB~\cite{Yin:2024:ThinkRepair:SelfDirectedAutomatedProgramRepair}, while
MORepair was evaluated on SWE-Repair (a subset of SWE-Bench~\cite{Jimenez:2024:SWEBench:CanLanguageModelsResolveRealWorldGithubIssues}), EvalRepair-C++, EvalRepair-Java~\cite{Yang:2026:MORepair:TeachingLLMstoRepairCodeviaMultiObjectiveFineTuning}.
Regardless, as we do not train an LLM and program transformations are used, data leakage has limited impact on \MT{}.

\subsubsection{Prediction collection}
The original evaluations of the studied systems have already performed data cleaning to avoid data leakage.
Therefore, we directly replicated their experiments using the original settings. 
Specifically, both ThinkRepair and ReinFix were evaluated on Defects4J and RWB datasets, producing 556 and 720 repairs.
MORepair was evaluated on Defects4J, EvalRepair-C++, and EvalRepair-Java, and SWE-Repair, producing 902 repairs.

\subsubsection{Data cleaning}
We excluded incomplete responses, resulting in 432, 711, and 615 completed responses from ThinkRepair, ReinFix, and MORepair, respectively.
We further removed duplicates based on their MD5 hash code.
Ultimately, we curated 1,756 responses of patches and explanations, including 431, 711, and 614 from ThinkRepair, ReinFix, and MORepair, respectively.

\subsubsection{Data Labeling}

In this step, we labeled the correctness of 1,756 APR explanations.
Two annotators, each with 3--5 years of experience in programming and program repair, were assigned to independently label them.
For each explanation, the annotators were given the buggy program, the developer's patch, the LLM's explanation, the LLM's patch, and the auxiliary information such as the bug report and triggering tests.
They were instructed to review these artifacts to understand the bug.
Then, they judged whether the explanation is correct and indicated their confidence on a five-point Likert scale.
If one of them was unsure or the two annotators disagreed, a third annotator with over five years of relevant experience participated in the discussion to reach a consensus.
Details about the labeling process are described in Supplementary Material~\ref{sec:AppendixManualLabel}.
Table~\ref{table:overall_benchmark} shows the final results.

\begin{table}[t]
\caption{Benchmarking explanations generated by LLM-based APR systems}
\label{table:overall_benchmark}
\begin{tabular}{l|r|r|r|r}
\toprule
\textbf{APR System} & \textbf{\# Expls} & \textbf{\# Correct} & \textbf{\# Incorrect} & \textbf{Incorrect} \\
\textbf{Under Test} &  & \textbf{Expls} & \textbf{Expls} & \textbf{Expl Rate} \\
\midrule
ThinkRepair~\cite{Yin:2024:ThinkRepair:SelfDirectedAutomatedProgramRepair} & 431 & 54 & 377 & 87.5\% \\
ReInFix~\cite{Zhang:2026:RepairIngredientsAreAllYouNeed:ImprovingLargeLanguageModelBasedProgramRepairviaRepairIngredientsSearch} & 711 & 305 & 406 & 57.1\% \\
MORepair~\cite{Yang:2026:MORepair:TeachingLLMstoRepairCodeviaMultiObjectiveFineTuning} & 614 & 180 & 434 & 70.7\% \\
\midrule
Total & 1,756 & 539 & 1,217 & 69.3\% \\
\bottomrule
\end{tabular}
\end{table}

%% file: sections/results.tex
\input{tables/all}

\subsection{RQ1: Benchmark}

The annotation process achieves a Cohen's Kappa value~\cite{TheKappaStatistic} for inter-annotator agreement of 0.78, indicating substantial consistency and reliability.
This has significantly bolstered our confidence in both the benchmark's quality and the evaluation of the detection approach, \MT.

Table~\ref{table:overall_benchmark} shows the results of the labeling process.
It is clear that many generated explanations are incorrect. 
Among 1,756 explanations produced, 1,217 are incorrect, accounting for 69.3\% of all explanations.
This shows that the systems are still highly unreliable for explaining their repairs, even the best-performing agentic system, ReInFix.
A breakdown of the benchmark is shown in Supplementary Material~\ref{sec:AppendixResults}.

\begin{custombox}[RQ1]
All APR systems frequently generate incorrect explanations, with an overall incorrect rate of \textbf{69.3\%}.
\end{custombox}

\subsection{RQ2: Detection Effectiveness of \MT}

In RQ2, we examine whether \MT{} can achieve better effectiveness in detecting correct explanations than the baselines.
Table~\ref{table:main_result} compares their effectiveness across three LLM-based APR systems: ThinkRepair, ReInFix, and MORepair.

Overall, \textbf{the baselines achieve poor effectiveness}.
SelfCheckGPT is the strongest baseline on ReInFix and MORepair in terms of MCC (0.17 and 0.09, respectively).
The main issue is its low Recall of 0.09--0.38, despite high Specificity of 0.69--0.99. 
This means SelfCheckGPT misclassifies many incorrect explanations as correct. 
MetaQA is generally unstable and close to random. 
It achieves MCC of -0.02--0.01 on ThinkRepair and MORepair, only getting better MCC of 0.16 on ReInFix.
MetaQA's Recall is low (0.27–0.48), showing that it misses many incorrect explanations. 
DrHall is the poorest baseline, with MCC of -0.06--0.12.

Adding program transformations into SelfCheckGPT, as expected, improves Recall by 97.4\%--677.8\% but severely hurts Specificity by 47.1\%--92.6\%. 
This causes MCC to become negative on ReInFix (-0.07) and MORepair (-0.14). 
SelfCheck w/ ProgTrans tends to over-detect explanations as incorrect, failing to identify many correct ones. 


In contrast, \textbf{\MT{} consistently achieves strong effectiveness} across the APR systems.
It achieves 0.34--0.43 in MCC and 0.70--0.75 in BA.
Compared with the best baseline on each APR system, \MT{} improves MCC substantially by up to 580.0\%. 
\MT{} maintains relatively higher Precision of 0.88–0.96 and much higher Recall of 0.50--0.72. 
These results contrast with SelfCheckGPT and MetaQA, with lower Precision and Recall, and DrHall, which can achieve high Precision and Recall by always classifying explanations as incorrect.
Therefore, \MT{} provides a better balance between detecting incorrect explanations.


\begin{custombox}[RQ2]
\MT{} consistently outperforms the baselines in detection quality across all APR systems and bug datasets.
It achieves an improvement of up to \textbf{333.3\%} in MCC over SelfCheckGPT, the best detection baseline.
\end{custombox}

\subsection{RQ3: Token Usage}

Since \MT{} involves multiple rounds of LLM queries, it may consume more tokens than direct LLM prompts.
We quantify this increase and investigate the trade-off between token use and detection quality.
Table~\ref{table:main_result} provides the token usage across detectors and APR systems.
For each detector, we measure the input tokens, the output tokens, and total tokens.

SelfCheckGPT and SelfCheck~w/~ProgTrans consume tokens comparable to \MT{} on ThinkRepair and MORepair.
On ReInFix, they consume 0.48 times more tokens on average than \MT{}, due to the design of ReInFix's query.
\MT{} consumes 0.92 times more tokens than MetaQA and 25.79 times more tokens than DrHall, two more direct prompting methods.
These results reveal a clear trade-off between computational cost and detection effectiveness.
\MT{} consumes comparable or more tokens than most baselines, but justifies that cost by superior detection effectiveness.


\begin{custombox}[RQ3]
\MT{} consumes similar or even fewer tokens than SelfCheckGPT by 0.39 times.
It consumes 0.92 times more tokens than MetaQA and 25.79 times more tokens than DrHall.
\MT's superior performance boost sometimes comes at the cost of increased token usage.
\end{custombox}

\subsection{RQ4: Sensitivity Analysis}

\begin{figure*}
    \centering
    \includegraphics[scale=0.54, clip, trim=3mm 0mm 2mm 3mm]{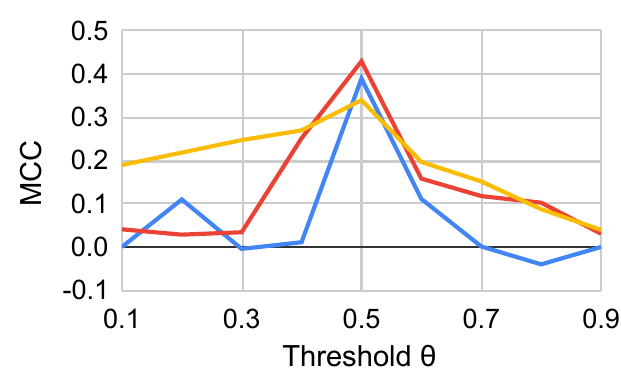}
    \hfill
    \includegraphics[scale=0.54, clip, trim=3mm 0mm 2mm 3mm]{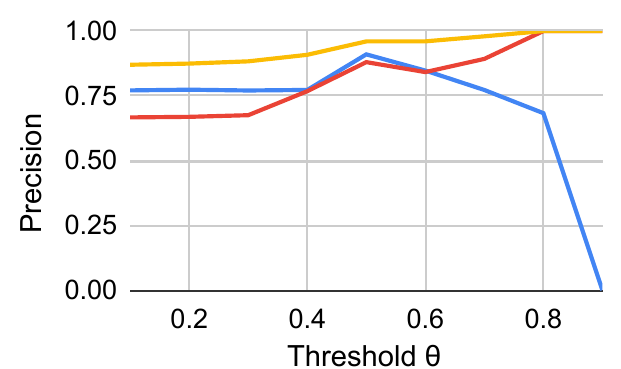}
    \hfill
    \includegraphics[scale=0.54, clip, trim=3mm 0mm 2mm 3mm]{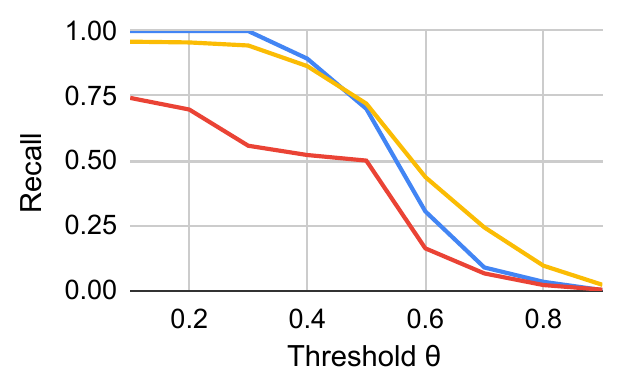}
    \caption{\MT's effectiveness with different values of $\theta$ across three APR systems: \colorbox[RGB]{251,188,4}{\makebox(0,5){}}~ThinkRepair, \colorbox[RGB]{234,67,53}{\makebox(0,5){}}~ReInFix, and \colorbox[RGB]{66,133,244}{\makebox(0,5){}}~MORepair.}
    \label{fig:sensitivity_theta}
\end{figure*}
\begin{figure*}
    \centering
    \includegraphics[scale=0.54, clip, trim=3mm 0mm 2mm 3mm]{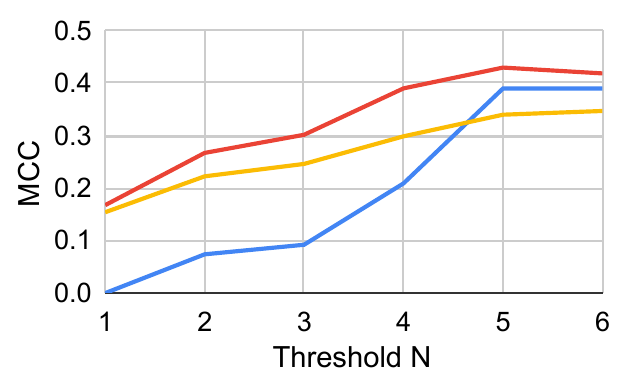}
    \hfill
    \includegraphics[scale=0.54, clip, trim=3mm 0mm 2mm 3mm]{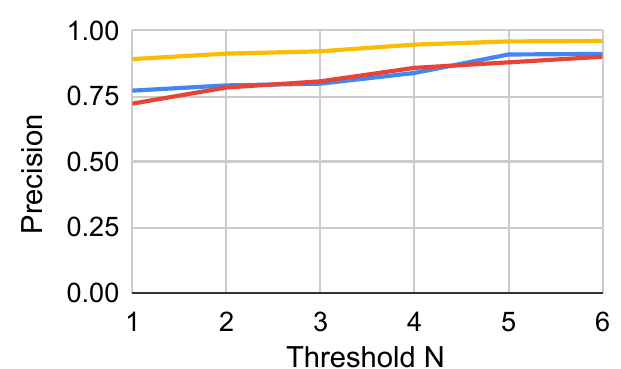}
    \hfill
    \includegraphics[scale=0.54, clip, trim=3mm 0mm 2mm 3mm]{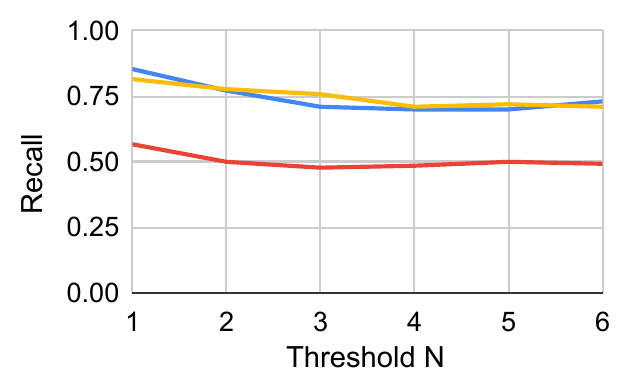}
    \caption{\MT's effectiveness with different values of $N$ across three APR systems: \colorbox[RGB]{251,188,4}{\makebox(0,5){}}~ThinkRepair, \colorbox[RGB]{234,67,53}{\makebox(0,5){}}~ReInFix, \colorbox[RGB]{66,133,244}{\makebox(0,5){}}~MORepair.}
    \label{fig:sensitivity_N}
\end{figure*}
RQ4 further investigates the sensitivity of \MT{} to two thresholds, $\theta$ and $N$, across the APR systems. 
Figure~\ref{fig:sensitivity_theta} shows \MT's effectiveness with different values of $\theta$.
Across all APR systems, \MT{} achieves the best MCC consistently around $\theta$$=$$0.5$. 
When $\theta$ is small, Precision remains low, while Recall is high.
This indicates that \MT{} relaxes the detection condition and tends to detect explanations as incorrect. 
As $\theta$ increases, Precision improves rapidly while Recall decreases.
However, after $\theta$$=$$0.5$, Recall drops substantially, meaning that more incorrect explanations are misclassified as correct. 
This degradation is reflected in MCC, which decreases after $\theta$$=$$0.5$ for all systems. 
These results suggest that $\theta$$=$$0.5$ provides the best detection for incorrect explanations.

Figure~\ref{fig:sensitivity_N} shows \MT's effectiveness with different values of $N$.
Increasing $N$ generally improves MCC and Precision across all three APR systems, while Recall changes only slightly. 
This indicates that sampling more explanations helps \MT{} detect incorrect explanations more effectively. 
The gains are most evident from $N=1$ to $N=5$.
After $N=5$, \MT's effectiveness becomes more stable. 


\begin{custombox}[RQ4]
\MT{} shows a clear trade-off between Precision and Recall as $\theta$ changes: increasing $\theta$ improves Precision but reduces Recall.
$\theta=0.5$ achieves the balance with the best MCC.
\MT's effectiveness is improved when increasing $N$ and becomes more stable when $N \geq 5$.
\end{custombox}

\subsection{RQ5: Ablation Study}

We conduct an ablation study to understand the contribution of \MT's three components: identifying contradictions by the same LLM, the equivalent questioning, and non-equivalent questioning.
We develop three variants of \MT{}.
The first one relies on an external LLM to identify contradictions instead of using the same LLM.
Claude Opus 4.5 (claude-opus-4-5-20251101) is selected as it is the best LLM for SE and is different from the LLM families used in APR systems.
The next two variants only use the equivalent questioning or non-equivalent questioning, respectively.
The results show that all components are necessary for \MT's detection quality. 
Removing a component consistently degrades MCC and balanced accuracy across all three APR systems. 

When \MT{} relies on Claude Opus 4.5 to identify contradictions, its MCC decreases by up to 32.6\%.
This variant often achieves slightly higher Recall of 0.75--0.79, except for lower Recall of 0.48 on ReInFix.
It achieves lower Precision of 0.76--0.92 (drop by up to 13.6\%) and lower Specificity of 0.60--0.80 (drop by up to 22.1\%).
Although its Recall is slightly better, many explanations are misclassified, leading to the overall decrease in MCC.

Compared to \MT{}, removing the non-equivalent questioning reduces MCC by up to 84.6\%.
This variant often achieves high Specificity of 0.80--0.93, but at the cost of a much lower Recall of 0.11--0.31.
This indicates that it tends to misclassify incorrect explanations as correct. 

In contrast, removing the equivalent questioning decreases MCC by up to 94.9\%.
Doing so substantially harms the ability to differentiate incorrect and correct explanations. 
This variant often has lower Specificity, especially on MORepair, where Specificity drops to 0.09. 
Although it sometimes achieves higher Recall, the big drops in Specificity show that it becomes overly relaxed and often mislabels explanations as incorrect.

\begin{custombox}[RQ5]
When using an external LLM to identify contradictions, \MT{} maintains slightly higher Recall but achieves lower Precision.
This leads to the overall decrease in MCC of up to \textbf{32.6\%}.
Additionally, both equivalent and non-equivalent questioning in \MT{} are crucial.
Removing them reduces \MT's detection ability by up to \textbf{84.6\%} and \textbf{94.9\%} in MCC, respectively.
\end{custombox}

\subsection{Qualitative Error Analysis}
To better understand the limitations of \MT{}, we manually analyzed a subset of misclassifications. 
The most common reason for incorrect explanations being misdetected is that the LLM is stubborn and consistently defends its wrong reasoning under \MT.
We provide a few examples in Supplementary Material~\ref{sec:AppendixExamples}.
This is a well-known limitation of sampling-based detection methods. 
Future work can explore different sampling techniques to better reveal LLMs' contradictions.
On the other hand, correct explanations misclassified as incorrect are rare. 
As shown in Supplementary Material~\ref{sec:AppendixProof}, there are only a few outliers, where the LLM generates a correct explanation but incorrectly verifies the follow-up explanations.

%% file: tables/all.tex
\begin{table*}[t]
\caption{The effectiveness in detecting incorrect APR explanations of \MT{} and the other methods.}
\label{table:main_result}
\centering
\renewcommand{\arraystretch}{1.1}
\begin{tabular}{|c|c|l|rrrrrr|rrr|}
\hline
\textbf{APR System} & \multirow{2}{*}{\textbf{\# Expls}} & \multirow{2}{*}{\textbf{Detector}} & \multicolumn{6}{c|}{\textbf{Detection Quality}} & \multicolumn{3}{c|}{\textbf{Tokens (Millions)}} \\
\cline{4-12}
\textbf{Under Test} & & & \textbf{MCC}$\uparrow$ & \textbf{$\Delta$MCC}$\uparrow$ & \textbf{BA}$\uparrow$ & \textbf{Pre}$\uparrow$ & \textbf{Rec}$\uparrow$ & \textbf{Spe}$\uparrow$ & \textbf{Input} & \textbf{Output} & \textbf{Total} \\
\hline \hline
\multirow{8}{*}{ThinkRepair~\cite{Yin:2024:ThinkRepair:SelfDirectedAutomatedProgramRepair}} & \multirow{8}{*}{431} & SelfCheckGPT~\cite{Manakul:2023:SelfCheckGPT:ZeroResourceBlackBoxHallucinationDetectionforGenerativeLargeLanguageModels} & 0.05$^{*}$ & \textcolor{red}{-50.0\%} & 0.53$^{*}$ & 0.89$^{*}$ & 0.38$^{*}$ & 0.68$^{\diamond}$ & 8.96 & 4.89 & 13.85 \\
& & SelfCheck w/ ProgTrans & 0.10$^{*}$ & 0\% & 0.56$^{*}$ & 0.89$^{*}$ & 0.75$^{\phantom{*}}$ & 0.36$^{*}$ & 9.15 & 4.97 & 14.12 \\
& & MetaQA~\cite{Yang:2025:HallucinationDetectioninLargeLanguageModelswithMetamorphicRelations} & -0.02$^{*}$ & \textcolor{red}{-120.0\%} & 0.49$^{*}$ & 0.87$^{*}$ & 0.37$^{*}$ & 0.61$^{*}$ & 4.06 & 2.55 & 6.61 \\
& & DrHall~\cite{Wu:2025:DetectingandReducingtheFactualHallucinationsofLargeLanguageModelswithMetamorphicTesting} & -0.06$^{*}$ & \textcolor{red}{-160.0\%} & 0.49$^{*}$ & 0.87$^{*}$ & 0.98$^{*}$ & 0.00$^{*}$ & 0.29 & 0.17 & 0.46 \\
\cline{3-12}
& & \textbf{\MT{} (ours)} & \textbf{0.34}$^{\phantom{*}}$ & \textbf{\color{pgreen}+240.0\%} & \textbf{0.75}$^{\phantom{*}}$ & 0.96$^{\phantom{*}}$ & 0.72$^{\phantom{*}}$ & 0.77$^{\phantom{*}}$ & 9.19 & 5.20 & 14.39 \\
\cline{3-12}
& & \MT{} w/ ext. Claude & 0.25$^{*}$ & \textcolor{pgreen}{+150.0\%} & 0.68$^{*}$ & 0.92$^{\phantom{*}}$ & 0.75$^{\diamond}$ & 0.60$^{*}$ & 9.30 & 5.11 & 14.41 \\
& & \MT{} w/o non-eqv. & 0.08$^{\ast}$ & \textcolor{red}{-20.0\%} & 0.56$^{*}$ & 0.92$^{\diamond}$ & 0.31$^{*}$ & 0.80$^{*}$ & 4.54 & 2.57 & 7.11 \\
& & \MT{} w/o eqv. & 0.23$^{\ast}$ & \textcolor{pgreen}{+130.0\%} & 0.64$^{*}$ & 0.91$^{*}$ & 0.83$^{*}$ & 0.45$^{*}$ & 4.65 & 2.63 & 7.28 \\
\hline \hline
\multirow{8}{*}{ReInFix~\cite{Zhang:2026:RepairIngredientsAreAllYouNeed:ImprovingLargeLanguageModelBasedProgramRepairviaRepairIngredientsSearch}} & \multirow{8}{*}{711} & SelfCheckGPT~\cite{Manakul:2023:SelfCheckGPT:ZeroResourceBlackBoxHallucinationDetectionforGenerativeLargeLanguageModels} & 0.17$^{*}$ & 0\% & 0.54$^{*}$ & 0.92$^{*}$ & 0.09$^{*}$ & 0.99$^{\diamond}$ & 30.41 & 2.52 & 32.93 \\
& & SelfCheck w/ ProgTrans & -0.07$^{*}$ & \textcolor{red}{-141.2\%} & 0.47$^{*}$ & 0.55$^{*}$ & 0.70$^{*}$ & 0.24$^{*}$ & 32.58 & 2.45 & 35.03 \\
& & MetaQA~\cite{Yang:2025:HallucinationDetectioninLargeLanguageModelswithMetamorphicRelations} & 0.16$^{*}$ & \textcolor{red}{-5.9\%} & 0.57$^{*}$ & 0.72$^{*}$ & 0.27$^{*}$ & 0.86$^{\diamond}$ & 7.43 & 4.67 & 12.1 \\
& & DrHall~\cite{Wu:2025:DetectingandReducingtheFactualHallucinationsofLargeLanguageModelswithMetamorphicTesting} & 0.12$^{*}$ & \textcolor{red}{-29.4\%} & 0.54$^{*}$ & 0.72$^{*}$ & 0.19$^{*}$ & 0.90$^{*}$ & 0.52 & 0.28 & 0.80 \\
\cline{3-12}
& & \textbf{\MT{} (ours)} & \textbf{0.43}$^{\phantom{*}}$ & \textbf{\textcolor{pgreen}{+152.9\%}} & \textbf{0.70}$^{\phantom{*}}$ & 0.88$^{\phantom{*}}$ & 0.50$^{\phantom{*}}$ & 0.91$^{\phantom{*}}$ & 15.08 & 8.60 & 23.68 \\
\cline{3-12}
& & \MT{} w/ ext. Claude & 0.29$^{*}$ & \textcolor{pgreen}{+70.6\%} & 0.64$^{*}$ & 0.76$^{*}$ & 0.48$^{\phantom{*}}$ & 0.80$^{*}$ & 15.12 & 7.35 & 22.47 \\
& & \MT{} w/o non-eqv. & 0.20$^{*}$ & \textcolor{pgreen}{+17.6\%} & 0.57$^{*}$ & 0.79$^{*}$ & 0.23$^{*}$ & 0.92$^{\phantom{*}}$ & 7.59 & 4.11 & 11.70 \\
& & \MT{} w/o eqv. & 0.18$^{*}$ & \textcolor{pgreen}{+5.9\%} & 0.58$^{*}$ & 0.63$^{*}$ & 0.77$^{*}$ & 0.40$^{*}$ & 7.49 & 4.49 & 11.98 \\
\hline \hline
\multirow{8}{*}{MORepair~\cite{Yang:2026:MORepair:TeachingLLMstoRepairCodeviaMultiObjectiveFineTuning}} & \multirow{8}{*}{614} & SelfCheckGPT~\cite{Manakul:2023:SelfCheckGPT:ZeroResourceBlackBoxHallucinationDetectionforGenerativeLargeLanguageModels} & 0.09$^{*}$ & 0\% & 0.53$^{*}$ & 0.84$^{*}$ & 0.11$^{*}$ & 0.95$^{*}$ & 9.31 & 5.12 & 14.43 \\
& & SelfCheck w/ ProgTrans & -0.14$^{*}$ & \textcolor{red}{-255.6\%} & 0.45$^{*}$ & 0.68$^{*}$ & 0.82$^{*}$ & 0.07$^{*}$ & 9.39 & 5.25  & 14.64 \\
& & MetaQA~\cite{Yang:2025:HallucinationDetectioninLargeLanguageModelswithMetamorphicRelations} & -0.01$^{*}$ & \textcolor{red}{-111.1\%} & 0.50$^{*}$ & 0.70$^{*}$ & 0.48$^{*}$ & 0.51$^{*}$ & 5.15 & 5.86 & 11.01 \\
& & DrHall~\cite{Wu:2025:DetectingandReducingtheFactualHallucinationsofLargeLanguageModelswithMetamorphicTesting} & 0.08$^{*}$ & \textcolor{red}{-11.1} & 0.54$^{*}$ & 0.75$^{*}$ & 0.45$^{*}$ & 0.64$^{\diamond}$ & 0.33 & 0.59 & 0.92 \\
\cline{3-12}
 & & \textbf{\MT{} (ours)} & \textbf{0.39}$^{\phantom{*}}$ & \textbf{\textcolor{pgreen}{+333.3\%}} & \textbf{0.71}$^{\phantom{*}}$ & 0.91$^{\phantom{*}}$ & 0.70$^{\phantom{*}}$ & 0.72$^{\phantom{*}}$ & 9.44 & 8.50 & 17.94 \\
\cline{3-12}
& & \MT{} w/ ext. Claude & 0.36$^{\diamond}$ & \textcolor{pgreen}{+300.0\%} & 0.70$^{\phantom{*}}$ & 0.83$^{*}$ & 0.79$^{\diamond}$ & 0.60$^{*}$ & 10.34 & 7.41 & 17.75 \\
& & \MT{} w/o non-eqv. & 0.06$^{*}$ & \textcolor{red}{--33.3\%} & 0.52$^{*}$ & 0.79$^{*}$ & 0.11$^{*}$ & 0.93$^{*}$ & 4.24 & 3.85 & 8.09 \\
& & \MT{} w/o eqv. & 0.02$^{*}$ & \textcolor{red}{-77.8\%} & 0.50$^{*}$ & 0.71$^{*}$ & 0.92$^{*}$ & 0.09$^{*}$ & 5.20 & 4.65 & 9.85 \\
\hline
\end{tabular}

\vspace{4pt}
{\footnotesize Abbreviations: MCC = Matthews Correlation Coefficient, $\Delta$MCC: Relative MCC Improvement Over the Best Baseline, BA = Balanced Accuracy, \\
Pre = Precision, Rec = Recall, Spe = Specificity. Paired t-test between \MT{} and other methods, with $\diamond$: p-value $<$ 0.05, $*$: p-value $<$ 0.01.}
\end{table*}

%% file: sections/threats.tex
\subsubsection*{Construct Validity}

We constructed a new benchmark based on recent APR studies.
They have already mitigated data leakage.
In the benchmark, each record is manually labeled by at least two annotators with 3--5 years of programming and repair experience.
The labeling process achieved a Cohen's Kappa of 0.78 for inter-annotator agreement. 
This suggests substantial consistency and reliability of labels. 
We acknowledge that the labeling is imperfect, leaving a small portion of mislabeled data. 
However, we believe the results still hold good.

\subsubsection*{Internal Validity}
LLMs are inherently nondeterministic, leading to varied responses. 
To address this, we set the temperature to 0 and employed three runs of experiments. 
Another threat is data leakage.
Although we do not train an LLM and program transformations are used, data leakage may have limited impact on \MT{}.
We mitigate this threat by using the experimental design in prior studies~\cite{Yin:2024:ThinkRepair:SelfDirectedAutomatedProgramRepair, Zhang:2026:RepairIngredientsAreAllYouNeed:ImprovingLargeLanguageModelBasedProgramRepairviaRepairIngredientsSearch, Yang:2026:MORepair:TeachingLLMstoRepairCodeviaMultiObjectiveFineTuning} of LLMs and datasets.
They have already mitigated data leakage.

\subsubsection*{External Validity}
Main threats to external validity are that the selected dataset may not generalize to other programs, or that performance enhancements will not generalize to other LLMs. 
While the benchmark contains three different programming languages (Java, Python, and C++), future studies on alternative datasets could further validate generalizability and performance trends. 
We evaluated \MT{} across three recent SOTA LLM-based APR systems, including both open-source and commercial LLMs. 
This broad spectrum enhances generalizability; however, different systems may yield varying results. 
Nevertheless, observed trends suggest broad applicability.
Evaluating \MT{} on larger and more powerful models may help assess its scalability and effectiveness.

\subsubsection*{Conclusion Validity} 
We interpreted the results using multiple well-established metrics: Precision, Recall, Specificity, and MCC. 
We also used appropriate statistical tests, such as the paired t-test, to support our findings.

%% file: sections/related_work.tex

\subsection{Automated Program Repair with LLMs}

Novel approaches have been designed to maximize the repair potential of LLMs.
Many approaches can generate correct patches for buggy programs with a success rate of up to 70\%~\cite{Zhao:2024:EnhancingAutomatedProgramRepairwithSolutionDesign, Le:2025:MemoryEfficientLargeLanguageModelsforProgramRepairwithSemanticGuidedPatchGeneration}.
In recent years, approaches have also been developed to explain what the bug is and how to fix it.

Many studies have adopted the fine-tuning paradigm~\cite{Yang:2026:MORepair:TeachingLLMstoRepairCodeviaMultiObjectiveFineTuning, Zhang:2024:VulAdvisor:NaturalLanguageSuggestionGenerationforSoftwareVulnerabilityRepair}.
This allows LLMs to learn empirical knowledge used to guide bug repair from bug-fix history. 
For example, MORepair~\cite{Yang:2026:MORepair:TeachingLLMstoRepairCodeviaMultiObjectiveFineTuning} leverages explanations of code changes to capture the logic underlying a repair operation during fine-tuning.

In addition, researchers have utilized the prompt-learning paradigm~\cite{Xia:2024:AutomatedProgramRepairviaConversation:Fixing162outof337Bugsfor0.42EachusingChatGPT, Yin:2024:ThinkRepair:SelfDirectedAutomatedProgramRepair} to stimulate the repair potential of LLMs.
They often combine LLMs with traditional APR techniques to construct specific prompt templates. 
For example, ThinkRepair~\cite{Yin:2024:ThinkRepair:SelfDirectedAutomatedProgramRepair} enables LLMs to better implement reasoning to complete the repair by leveraging CoT prompting~\cite{Wei:2022:ChainofThoughtPromptingElicitsReasoninginLargeLanguageModels}. 

Recent research has entered the era of agents.
Researchers have allowed LLMs to take on different task roles, further expanding the repair capabilities of LLMs. 
For example, 
AutoCodeRover~\cite{Zhang:2024:AutoCodeRover:AutonomousProgramImprovement} combines LLMs with sophisticated code search capabilities to generate patches.
RepairAgent~\cite{Bouzenia:2025:RepairAgent:AnAutonomousLLMBasedAgentforProgramRepair} enables the LLM-based agent to perform repair by invoking external tools.
ReInFix~\cite{Zhang:2026:RepairIngredientsAreAllYouNeed:ImprovingLargeLanguageModelBasedProgramRepairviaRepairIngredientsSearch} allows LLMs to search for repair ingredients, such as variables and historical changes, to better analyze bug causes and guide bug fixing.

\subsection{The Detection of Incorrect LLM Responses}

Incorrect responses generated by LLMs are often referred to as hallucinations.
Detecting them is essential for assuring the trustworthiness of LLM-generated content.
Many detection approaches have been introduced for both NL~\cite{Manakul:2023:SelfCheckGPT:ZeroResourceBlackBoxHallucinationDetectionforGenerativeLargeLanguageModels, Yang:2025:HallucinationDetectioninLargeLanguageModelswithMetamorphicRelations, Cohen:2023:LMvsLM:DetectingFactualErrorsviaCrossExamination, Lam:2025:TOKI} and code artifacts~\cite{Tian:2025:CodeHalu:InvestigatingCodeHallucinationsinLLMsviaExecutionbasedVerification, Yang:2026:AdvancingLLMGeneratedCodeReliability:AHybridApproachforHallucinationDetection, Agarwal:2025:CodeMirage:HallucinationsinCodeGeneratedbyLargeLanguageModels, Lam:2026:UntrustVul}.
A straightforward strategy is to compare the content against reliable knowledge sources~\cite{Guo:2022:ASurveyonAutomatedFactChecking}. 
However, this strategy requires curated databases, which are often unavailable and costly to construct.

To address this, prior work has explored zero-resource detection approaches.
One line of work exploits LLMs' internal states~\cite{Su:2024:UnsupervisedRealTimeHallucinationDetectionbasedontheInternalStatesofLargeLanguageModels, Du:2024:HaloScope:HarnessingUnlabeledLLMGenerationsforHallucinationDetection, Sriramanan:2024:LLMCheck:InvestigatingDetectionofHallucinationsinLargeLanguageModels, Kim:2025:DetectingLLMHallucinationThroughLayerWiseInformationDeficiency:AnalysisofAmbiguousPromptsandUnanswerableQuestions}
These methods do not require external knowledge and can be computationally efficient.
However, they assume access to the internal states, which are not always available.
A second line of work targets black-box settings.
Some methods adopt a multi-agent perspective by using additional LLMs to judge the original LLM's response~\cite{Wu:2025:DetectingandReducingtheFactualHallucinationsofLargeLanguageModelswithMetamorphicTesting, Cohen:2023:LMvsLM:DetectingFactualErrorsviaCrossExamination}.
Others use a sampling-based approach, which is often more practical~\cite{Yang:2025:HallucinationDetectioninLargeLanguageModelswithMetamorphicRelations}. 
They sample multiple responses from the LLM and identify if they contradict one another.
For instance, SelfCheckGPT~\cite{Manakul:2023:SelfCheckGPT:ZeroResourceBlackBoxHallucinationDetectionforGenerativeLargeLanguageModels} repeatedly queries the LLM and identifies contradictions. 
MetaQA~\cite{Yang:2025:HallucinationDetectioninLargeLanguageModelswithMetamorphicRelations} and DrHall~\cite{Wu:2025:DetectingandReducingtheFactualHallucinationsofLargeLanguageModelswithMetamorphicTesting} apply semantic perturbations and check if new responses remain valid.

The sampling-based methods have significantly advanced the detection of incorrect LLM-generated content, primarily in general question-answering. 
Unfortunately, these methods are not well-suited or effective for APR explanations.
This is because detection methods need APR-specific knowledge, and the correctness of APR explanations depends on the bug and the intended program behavior.
We introduce \MT{}, a more effective detection method of incorrect APR explanations by addressing these limitations.




%% file: sections/conclusion.tex
This work introduces \MT, the first approach to detect incorrect explanations of LLM-based APR systems. 
Inspired by debate and viva examination, \MT{} adopts a metamorphic argumentative strategy to identify contradictions in LLM responses.
\MT{} also applies sampling based on diverse program transformations to better reveal contradictions.
Our evaluation shows significant performance improvements in detection quality.
\MT{} improves MCC by up to 333.3\% over existing SOTA methods designed for general question-answering.
Although \MT{} sometimes consumes more tokens, it is an acceptable cost for improving the detection quality.
Future work can define more MRs to better detect incorrect APR explanations, as well as explore the application of \MT{} in other SE tasks.

%% file: sections/supplementary_material.tex
\twocolumn[
\begin{center}
{\Huge Supplemental Material of ``Detecting Incorrect LLM Explanations for Automated Program Repair''}
\end{center}
\vspace{8mm}
]

\section{Motivating Example}
\label{sec:Appendix_Example}

\begin{lstlisting}[language=java,label=lst:Program1,caption={Program1 semantically equivalent to \textcode{getTextBuffer}, as shown in Figure~\ref{fig:motivating_example}. It is generated by rewriting the program into nested if–else blocks.}]
public char[] getTextBuffer() {
    if (_inputStart >= 0) {
        return _inputBuffer;
    } else {
        if (_resultArray != null) {
            return _resultArray;
        } else {
            if (_resultString != null) {
                return _resultArray = _resultString.toCharArray();
            } else {
                if (!_hasSegments) {
                    return _currentSegment;
                } else {
                    return contentsAsArray();
                }
            }
        }
    }
}
\end{lstlisting}


\begin{lstlisting}[language=java,label=lst:Program3,caption={Program3 semantically non-equivalent to \textcode{getTextBuffer}, as shown in Figure~\ref{fig:motivating_example}. The comparison operator \textcode{>=} in \textcode{\_inputStart} \textcode{>=} \textcode{0} is changed to \textcode{>}.}]
public char[] getTextBuffer() {
    if (_inputStart > 0)
        return _inputBuffer;
    if (_resultArray != null)
        return _resultArray;
    if (_resultString != null) {
        return (_resultArray = _resultString.toCharArray());
    }
    if (!_hasSegments)
        return _currentSegment;
    return contentsAsArray();
}
\end{lstlisting}


\section{Detection Proof}
\label{sec:AppendixProof}

\subsubsection{Empirical Proof by Comparing MR Violation Rates}

\begin{figure}
    \centering
    \includegraphics[width=\linewidth, clip, trim=0 6mm 0 0]{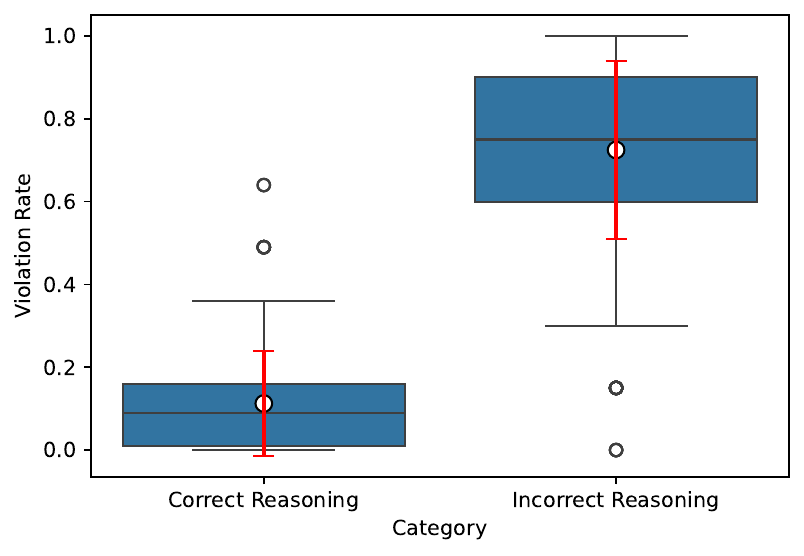}
    \caption{Comparing MR violation rates when the original LLM-generated explanations are correct and incorrect.}
    \label{fig:violation_rate_per_case}
\end{figure}

Figure~\ref{fig:violation_rate_per_case} shows a clear association between the correctness of the original LLM-generated explanation and the MR violation score.
Overall, the violation score across all cases is $0.54$.
For cases where the original explanation is correct, the violation score is $0.10 \pm 0.12$.
Most values fall close to 0.00, although there are some high outliers near 0.50--0.70.
For cases where the original explanation is incorrect, the violation score is $0.73 \pm 0.21$.
This is much higher than the violation score when the explanation is correct.
Most values fall close to 1.00.
There are a few low outliers, but the overall distribution is shifted upward.

\begin{figure}[t]
    \centering
    \includegraphics[width=0.56\linewidth]{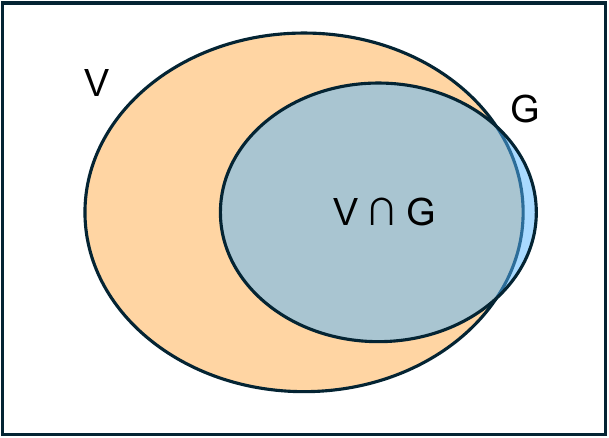}
    \caption{Venn diagram of correctly verifying explanations ($V$) and generating correct explanations ($G$).}
    \label{fig:venn_VG}
\end{figure}

Let $V$ represent that the LLM correctly verifies an explanation's correctness.
Let $G$ represent that the LLM generates a correct explanation.
Given the benchmark shown in Table~\ref{table:overall_benchmark}, we get $\mathbb{P}(\neg G) \approx 0.69$.
Based on the empirical results, we have $\mathbb{P}(\neg V) = 0.54$.
$\mathbb{P}(\neg V \mid G) \approx 0.10$, meaning $\mathbb{P}(V \mid G) \approx 0.90$.
$\mathbb{P}(\neg V \mid \neg G) \approx 0.73$, meaning $\mathbb{P}(V \mid \neg G) \approx 0.27$.
We can calculate $\mathbb{P}(G \mid \neg V)$ as follows
\begin{equation}
\begin{split}
\mathbb{P}(G \mid \neg V) & = \frac{\mathbb{P}(G \cap \neg V)}{\mathbb{P}(\neg V)} \\
& = \frac{\mathbb{P}(\neg V \mid G) \cdot \mathbb{P}(G)}{\mathbb{P}(\neg V)} \\
& =  \frac{\mathbb{P}(\neg V \mid G) \cdot \big( 1 -\mathbb{P}(\neg G) \big)}{\mathbb{P}(\neg V)} \\
& = \frac{0.10 \cdot (1 - 0.69)}{0.54} \approx 0.05
\end{split}
\end{equation}
$\mathbb{P}(V \mid G) \approx 0.90$ (close to 1), and $\mathbb{P}(V \mid \neg G) \approx 0.27 \gg \mathbb{P}(G \mid \neg V) \approx 0.05$.
Hence, $V$ is almost always no more difficult than $G$.
This relationship can be visualised in the Venn diagram in Figure~\ref{fig:venn_VG}.
Given $\neg V$, we can imply $\neg G$, with only a few false positives ($\neg V \cap G$).
On the other hand, false negatives ($V \cap \neg G$) are relatively more common.


\subsubsection{Theoretical Proof}

Let $m$ be the LLM-based APR system, $(r,y)$ be the original repair consisting of the explanation $r$ and the patch $y$.
Let $C(r) = \{0, 1\} $ indicate whether the original explanation $r$ is incorrect or correct.

We define $\text{\textcode{Tran}}_{\text{\textit{eqv}}}$ and $\text{\textcode{Tran}}_{\text{\textit{neq}}}$ for preserving and altering the semantics of a program, respectively.
We design a repair sub-task $\text{\textcode{Explain}}_m(x', y)$ that asks the system $m$ to explain how to fix transformed programs, explicitly given the original patch $y$.
We further model the verification task performed by $m$, where \MT{} asks $m$ to assess if a follow-up explanation applies to the original program.
We denote it by $\text{\textcode{Verify}}_m(r', x) = \{0, 1\}$, where the response of 1 indicates that the model $m$ thinks the explanation $r'$ can be applied to the program $x$, and 0 otherwise.

A violation of the two MRs can be represented as 
\[
\text{\textcode{Verify}}_m(r'_{\text{\textit{neq}}}, x) = 1 \lor \text{\textcode{Verify}}_m(r'_{\text{\textit{eqv}}}, x) = 0
\]
where $r'_{\text{\textit{eqv}}} \sim \text{\textcode{Explain}}_m(x'_{\text{\textit{eqv}}}, y) \circ \text{\textcode{Tran}}_{\text{\textit{eqv}}}(x)$ and $r'_{\text{\textit{neq}}} \sim \text{\textcode{Explain}}_m(x'_{\text{\textit{neq}}}, y) \circ \text{\textcode{Tran}}_{\text{\textit{neq}}}(x)$.

Our assumption is that when a system is capable of generating a correct explanation, it is also capable of performing easier follow-up tasks correctly.
\[C(r) = 1 \implies C(z) = 1 \,\text{, or }\, C(z) = 0 \implies C(r) = 0\] 
where $z \sim T_m(\cdot)$ and $T_m$ is less complex than $\text{\textcode{Gen}}_m$, i.e., $T_m = o\!\left(\text{\textcode{Gen}}_m\right)$.

$\text{\textcode{Explain}}_m$, and $\text{\textcode{Verify}}_m$ are designed to be easier than \textcode{Gen}$_m$.
A violation of the MRs represents failures in a sequence of tasks $\text{\textcode{Explain}}_m$, and $\text{\textcode{Verify}}_m$.
Hence, an MR violation indicates that \textcode{Gen}$_m$ fails.
In other words, 
\begin{align*}
\Big( \text{\textcode{Verify}}_m&(r'_{\text{\textit{neq}}}, x) = 1 \; \lor \\
& \text{\textcode{Verify}}_m(r'_{\text{\textit{eqv}}}, x) = 0 \Big) \implies C(r) = 0
\end{align*}
Taking corresponding probabilistic events
\[E_{\text{\textit{viol}},\text{\textit{eqv}}} := \{\text{\textcode{Verify}}_m(r'_{\text{\textit{eqv}}}, x) = 0\}\]
\[E_{\text{\textit{viol}},\text{\textit{neq}}} := \{\text{\textcode{Verify}}_m(r'_{\text{\textit{neq}}}, x) = 1\}\]
\[E_{Cr,0} := \{\text{\textcode{Verify}}_m(r'_{\text{\textit{eqv}}}, x) = 0\}\]
We thus have
\[E_{\text{\textit{viol}},\text{\textit{eqv}}}  \cup E_{\text{\textit{viol}},\text{\textit{neq}}}  \subseteq E_{Cr,0}\]
Therefore
\[\mathbb{P}(E_{\text{\textit{viol}},\text{\textit{eqv}}}  \cup E_{\text{\textit{viol}},\text{\textit{neq}}}) \leq \mathbb{P}(E_{Cr,0})\]
Since by definition, the violation score $\mathcal{V}_m(r, x) = \mathbb{P}(E_{\text{\textit{viol}},\text{\textit{eqv}}}  \cup E_{\text{\textit{viol}},\text{\textit{neq}}})$. 
We thus obtain
\[\mathcal{V}_m(r, x) \leq \mathbb{P}(E_{Cr,0})\]
The violation score acts as a lower bound on the incorrectness of the explanation $r$.

\section{Literature Survey of Program Transformation}
\label{sec:AppendixTranSurvey}

\subsubsection{Scope}
Our literature survey covers research published between 2021 and 2025, focusing on the most influential venues in software engineering. 
We selected conferences and journals known for their prominence and impact in the field. 
Specifically, we included the International Conference on Software Engineering (ICSE), Foundations of Software Engineering (FSE), International Symposium on Software Testing and Analysis (ISSTA), Automated Software Engineering (ASE), and Programming Language Design and Implementation (PLDI).
We also considered some leading venues in security and machine learning, such as Symposium on Security and Privacy (S\&P), Computer and Communications Security (CCS), USENIX Security Symposium (USENIX), International Conference on Machine Learning (ICML), Annual Meeting of the Association for Computational Linguistics (ACL), and Annual AAAI Conference on Artificial Intelligence (AAAI).
In addition, we incorporated top software engineering journals, such as Transactions on Software Engineering and Methodology (TOSEM), Transactions on Software Engineering (TSE), Journal of Systems and Software (JSS), Information and Software Technology (IST), and Empirical Software Engineering (ESE).
These venues were chosen for their reputation in publishing high-quality and influential research in software engineering (CORE\footnote{https://www.core.edu.au/home} rank A* or A).

\subsubsection{Criteria for Paper Selection}
To identify relevant publications, we established clear selection criteria. 
Eligible papers must:
\begin{itemize}
    \item Be full-length papers published in the proceedings or issues of the listed in-scope venues between 2021 and 2025, 
    \item Exclude short papers (e.g., PhD symposium or work-in-progress papers), as these typically represent incomplete research, and
    \item Address program transformations, meaning they either propose a new transformation technique or provide a review of existing ones.
\end{itemize}
To support paper identification, we defined a set of keywords that capture the research focus of this area: program, code, transformation, transform, modify, modification, perturb, perturbation, mutate, mutation, mutant, counterexample, adversarial, attack, semantic, preserve, alter, equivalence, and morph.
We searched for these keywords in the titles and abstracts of all in-scope papers using multiple tools, including IEEE Xplore Advanced Search for IEEE venues, ACM DL Advanced Search for ACM venues, and Google Scholar Advanced Search for independent venues.
This process yielded 88 potential papers, which were flagged for further manual review.

\subsubsection{Filtering Process} 
We manually reviewed the flagged papers to exclude those not directly related to program transformations. 
A total of 49 papers were excluded as off-topic.
The remaining 39 papers are shown in Table~\ref{table:paper_pt}.

\begin{table*}[]
\caption{Papers about Program Transformation}
\label{table:paper_pt}
\centering
\setlength{\tabcolsep}{4pt} 
\resizebox{\textwidth}{!}{%
\begin{tabular}{p{16.5cm}|r|l}
\toprule
\textbf{Title} & \textbf{Year} & \textbf{Venue} \\
\midrule
Assessing Robustness of ML-Based Program Analysis Tools using Metamorphic Program Transformations & 2021 & ASE \\
On the generalizability of Neural Program Models with respect to semantic-preserving program transformations & 2021 & IST \\
Self-Supervised Contrastive Learning for Code Retrieval and Summarization via Semantic-Preserving Transformations & 2021 & SIGIR \\
Adversarial Robustness of Deep Code Comment Generation & 2022 & TOSEM \\
Bridging Pre-trained Models and Downstream Tasks for Source Code Understanding & 2022 & ICSE \\
CODE-MVP: Learning to Represent Source Code from Multiple Views with Contrastive Pre-Training & 2022 & NAACL \\
Data augmentation by program transformation & 2022 & JSS \\
NatGen: generative pre-training by “naturalizing” source code & 2022 & FSE \\
ReACC: A Retrieval-Augmented Code Completion Framework & 2022 & ACL \\
Selfapr: Self-supervised program repair with test execution diagnostics & 2022 & ASE \\
Towards Learning (Dis)-Similarity of Source Code from Program Contrasts & 2022 & ACL \\
Towards Robustness of Deep Program Processing Models—Detection, Estimation, and Enhancement & 2022 & TOSEM \\
TransRepair: Context-aware Program Repair for Compilation Errors & 2022 & ASE \\
A Black-Box Attack on Code Models via Representation Nearest Neighbor Search & 2023 & EMNLP \\
An Extensive Study on Adversarial Attack against Pre-trained Models of Code & 2023 & FSE \\
CCTEST: Testing and Repairing Code Completion Systems & 2023 & ICSE \\
Challenging Machine Learning-Based Clone Detectors via Semantic-Preserving Code Transformations & 2023 & TSE \\
Code Difference Guided Adversarial Example Generation for Deep Code Models & 2023 & ASE \\
CodeAttack: Code-Based Adversarial Attacks for Pre-trained Programming Language Models & 2023 & AAAI \\
CodeMark: Imperceptible Watermarking for Code Datasets against Neural Code Completion Models & 2023 & FSE \\
DIP: Dead code insertion based black-box attack for programming language model & 2023 & ACL \\
Discrete Adversarial Attack to Models of Code & 2023 & PLDI \\
How Effective Are Neural Networks for Fixing Security Vulnerabilities & 2023 & ISSTA \\
Is Your Code Generated by ChatGPT Really Correct? Rigorous Evaluation of Large Language Models for Code Generation & 2023 & NeurIPS \\
Repair Is Nearly Generation: Multilingual Program Repair with LLMs & 2023 & AAAI \\
ALANCA: Active Learning Guided Adversarial Attacks for Code Comprehension on Diverse Pre-trained and Large Language Models & 2024 & SANER \\
An LLM-Assisted Easy-to-Trigger Backdoor Attack on Code Completion Models: Injecting Disguised Vulnerabilities against Strong Detection & 2024 & USENIX \\
Benchmarking Automated Program Repair: An Extensive Study on Both Real-World and Artificial Bugs & 2024 & ISSTA \\
Do large code models understand programming concepts? counterfactual analysis for code predicates & 2024 & ICML \\
Effective test generation using pre-trained Large Language Models and mutation testing & 2024 & IST \\
Mutation-based Consistency Testing for Evaluating the Code Understanding Capability of LLMs & 2024 & CAIN \\
Poison Attack and Poison Detection on Deep Source Code Processing Models & 2024 & TOSEM \\
Stealthy Backdoor Attack for Code Models & 2024 & TSE \\
AdverIntent-Agent: Adversarial Reasoning for Repair Based on Inferred Program Intent & 2025 & ISSTA \\
CodeMorph: Mitigating Data Leakage in Large Language Model Assessment & 2025 & ICSE \\
How Accurately Do Large Language Models Understand Code? & 2025 &  \\
Metamorphic-Based Many-Objective Distillation of LLMs for Code-related Tasks & 2025 & ICSE \\
Towards Reliable Evaluation of Neural Program Repair with Natural Robustness Testing & 2025 & TOSEM \\
Understanding the Robustness of Transformer-Based Code Intelligence via Code Transformation: Challenges and Opportunities & 2025 & TSE \\
\bottomrule
\end{tabular}%
}
\end{table*}

\begin{table*}[t]
\caption{Semantic preserving transformations used in this study}
\label{table:semantic_preserving_transformation}
\setlength{\tabcolsep}{3pt} 
\begin{subtable}[t]{0.496\textwidth}
\resizebox{\textwidth}{!}{
\begin{tabular}[t]{|c|p{3cm}|p{2.3cm}|p{2.3cm}|}
\hline
\textbf{Id} & \multicolumn{1}{|c|}{\textbf{Transformation}} & \multicolumn{1}{|c|}{\textbf{Original}} & \multicolumn{1}{|c|}{\textbf{Transformed}} \\
\hline
\multicolumn{4}{|c|}{Stylistic Transformations} \\
\hline
PCS-1       & Change Spacing Style & if (x == 0) \newline \{ \newline\} & if(x==0)\{\newline\} \\
\hline
PCS-2       & Change Naming Style & myNumber & my\_number \\
\hline
\multicolumn{4}{|c|}{Unused or Dead Code Transformations} \\
\hline
PUD-1       & Add Unused Statements/Dead Code                                              &                                       &                                \\
\hline
PUD-2       & Delete Unused Statement                                                     &                                       &                           \\
\hline
PUD-3       & Add Unused Parameters/Variable                                              &                                       &                                            \\
\hline
PUD-4       & Move Partial Code into Function & z=x+y & def func(x,y):\newline return x+y\newline z=func(x,y) \\
\hline
PUD-5       & Delete Unused Variable                                                      &                                       &   \\
\hline
\multicolumn{4}{|c|}{Expression Transformations}   \\
\hline
PEL-1       & Wrap with Logical NOT & x$<$y & !(y$>=$x) \\
\hline
PEL-2       & Split Compound Condition into Nested Ifs & if (cond1\newline\&\& cond2) \{\newline//Body\newline\} & if (cond1) \{\newline if (cond2) \{\newline//Body\newline\}\} \\
\hline
PEL-3       & Merge Variable Declarations & int a;\newline int b; & int a, b; \\
\hline
PEL-4       & Separate Declaration and Initialization & int i=0; & int i; i=0; \\
\hline
PEL-5       & Combine Declaration and Initialization & int i; i=0; & int i=0; \\
\hline
PEL-6       & Transform Relational Expression & a $<$ b & b $>$ a \\
\hline
PEL-7       & Transform Boolean Logic Equivalently & x == -1 \newline i == j \newline (a $||$ b) \&\& c & x + 1 == 0 \newline j == i \newline (a \&\& c) $||$ (b \&\& c) \\
\hline
PEL-8       & Convert Add/Subtract Assignment to Equal Assignment & x += 1 \newline y -= 2 & x = x + 1\newline y = y - 2 \\
\hline
PEL-9       & Divide Expression into Small Expressions with Intermediate Variables & x = a + b * c \newline\newline return 1 \newline\newline get().run() \newline f(g()) & temp = b * c\newline x = a + temp \newline ret = 1\newline return ret \newline i=get(); i.run() \newline v = g(); f(v) \\
\hline
PEL-10      & Swap Operands & a + b & b + a \\
\hline
PEL-11      & Combine Small Expressions into Expression & temp = b * c\newline x = a + temp \newline ret = 1\newline return ret \newline i=get(); i.run() \newline v = g(); f(v) & x = a + b * c \newline\newline return 1 \newline\newline get().run() \newline f(g()) \\
\hline
PEL-12      & Change Arithmetic Expression Equivalently & a - b + c & a - (b - c) \\
\hline
\end{tabular}%
}
\end{subtable}
\hfill
\begin{subtable}[t]{0.496\textwidth}
\resizebox{\textwidth}{!}{
\begin{tabular}[t]{|c|p{3cm}|p{2.3cm}|p{2.3cm}|}
\hline
\textbf{Id} & \multicolumn{1}{|c|}{\textbf{Transformation}} & \multicolumn{1}{|c|}{\textbf{Original}} & \multicolumn{1}{|c|}{\textbf{Transformed}} \\
\hline
PNL-1       & Modify Comment                                &                                       &                                          \\
\hline
PNL-2       & Add Comment                                   &                                       &                                          \\
\hline
PNL-3       & Remove Comment                                &                                       &                                          \\
\hline
PNL-4       & Add Print Statement                           &                                       &                                          \\
\hline
PNL-5       & Remove Print Statement                        &                                       &                                          \\
\hline
PNL-6       & Modify Print Message                          &                                       &                                        \\
\hline
\multicolumn{4}{|c|}{Control Flow Transformations}                                                                                                \\
\hline
PCL-1       & Convert For-Loop to While-Loop                 &                                       &                                          \\
\hline
PCL-2       & Convert While-Loop to For-Loop                 &                                       &                                          \\
\hline
PCL-3       & Convert If–Elseif to If–Else  & if (a $<$ 3)\{\newline//BodyA\newline\} else if (a $<$ 5)\{ //BodyB \newline \} & if (a $<$ 3)\{\newline//BodyA\newline\} else \{ \newline if (a $<$ 5) \newline //BodyB \newline \} \\
\hline
PCL-4       & Convert If–Else to If–Elseif & if (a $<$ 3)\{\newline//BodyA\newline\} else \{ \newline if (a $<$ 5) \newline //BodyB \newline \} & if (a $<$ 3)\{\newline//BodyA\newline\} else if (a $<$ 5)\{ //BodyB \newline \} \\
\hline
PCL-5       & Convert Switch–Case to If–Else                 &                                       &                                          \\
\hline
PCL-6       & Convert If–Else to Switch–Case                 &                                       &                                          \\
\hline
PCL-7       & Convert If–Continue to If–Else                 &                                       &                                          \\
\hline
PCL-8       & Swap If and Else Blocks & if (condition)\{\newline//BodyA\newline\} else \{\newline//BodyB\newline\} & if (!condition)\{\newline//BodyB\newline\} else \{\newline//BodyA\newline\} \\
\hline
PCL-9       & Swap Two Independent Statements                &                                       &                                          \\
\hline
PCL-10      & Convert If Statement to Conditional Expression & if (condition)\{\newline//BodyA\newline\} else \{\newline//BodyB\newline\} & condition ? BodyA : BodyB \\
\hline
PCL-11      & Convert Conditional Expression to If Statement & condition ? BodyA : BodyB & if (condition)\{\newline//BodyA\newline\} else \{\newline//BodyB\newline\} \\
\hline
PCL-12      & Add Else Clause                                &                                       &                                          \\
\hline
PCL-13      & Wrap Statement in If or Loop                   & statement & if (1)\newline statement \\
\hline
PCL-14      & Move Iterator Declaration into For-Loop        & int i; for (i; i $<$ n; i++) & for (int i = 0; i $<$ n; i++) \\
\hline
PCL-15      & Move Iterator Declaration out of For-Loop      & for (int i = 0; i $<$ n; i++) & int i; for (i; i $<$ n; i++) \\
\hline
\end{tabular}
}
\end{subtable}
\end{table*}
\begin{table*}[t]
\caption{Semantic altering transformations used in this study}
\label{table:semantic_altering_transformation}
\setlength{\tabcolsep}{3pt} 
\begin{subtable}[t]{0.496\textwidth}
\resizebox{\textwidth}{!}{
\begin{tabular}[t]{|c|p{3.5cm}|p{2.05cm}|p{2.05cm}|}
\hline
\textbf{Id} & \multicolumn{1}{|c|}{\textbf{Transformation}} & \multicolumn{1}{|c|}{\textbf{Original}} & \multicolumn{1}{|c|}{\textbf{Transformed}} \\
\hline
\multicolumn{4}{|c|}{Identifier Based Transformations}                                                                                         \\
\hline
AIB-1                & Rename Variable                             &                                       &                                          \\
\hline
\multicolumn{4}{|c|}{Control Flow Transformations}                                                                                             \\
\hline
ACL-1                & Misplace Return                             &                                       &                                          \\
\hline
ACL-2                & Swap Break and Continue                     &                                       &                                          \\
\hline
ACL-3                & Delete Exception Handler                    & except: pass                          & except: raise                            \\
\hline
ACL-4                & Swallow Exception                           & except: return                   & except: pass                             \\
\hline
ACL-5                & Delete Statement                            &                                       &                                          \\
\hline
ACL-6                & Wrap Statement with Condition               &                                       &                                          \\
\hline
ACL-7                & Unwrap Block                                &                                       &                                          \\
\hline
ACL-8                & Delete Block                                &                                       &                                          \\
\hline
ACL-9                & Insert Block                                &                                       &                                          \\
\hline
ACL-10               & Misuse Variable                             &                                       &                                          \\
\hline
\end{tabular}%
}
\end{subtable}
\hfill
\begin{subtable}[t]{0.496\textwidth}
\resizebox{\textwidth}{!}{
\begin{tabular}[t]{|c|p{3.5cm}|p{2.05cm}|p{2.05cm}|}
\hline
\textbf{Id} & \multicolumn{1}{|c|}{\textbf{Transformation}} & \multicolumn{1}{|c|}{\textbf{Original}} & \multicolumn{1}{|c|}{\textbf{Transformed}} \\
\hline
\multicolumn{4}{|c|}{Expression Transformations}                                                                                               \\
\hline
PEL-1       & Off-By-One                                  &  n                                     & n+1                                         \\
\hline
PEL-2       & Replace Conditional Expression              & a $||$ b                                & a                                        \\
\hline
PEL-3       & Expand Boolean Expression                   & x \&\& y                              & x \&\& y $||$ z                            \\
\hline
PEL-4       & Replace Relational Operator                 &   x $<$ 10                                    &  x $<=$ 10                                        \\
\hline
PEL-5       & Replace Boolean Operator                    &   x \&\& y                                    &  x $||$ y                                        \\
\hline
PEL-6       & Replace Literal Value                       &                                       &                                          \\
\hline
PEL-7      & Negative Condition                          &   !cond                                     &  cond                                  \\
\hline
PEL-8      & Remove Slice Index                          & sorted(l{[}::3{]})       & sorted(l{[}:{]})            \\
\hline
PEL-9      & Modify Function Call                        &                                       &                                          \\
\hline
\end{tabular}
}
\end{subtable}
\end{table*}

\section{Prompt Templates}
\label{sec:PromptTemplate}


\begin{lstlisting}[language=LLMPrompt, caption={Follow-up Explanation Generation}]
System: You are an Automatic Program Repair Tool. You are now the examinee in a debate-style viva examination, where you defend your original repair decision.

User: You originally provided the following fix for the original buggy function:
// Buggy Function
{{original_buggy_code}}

// Fixed Function
{{fixed_code}}

Now, the examiner presents a transformed buggy function:
// Buggy Function
{{transformed_buggy_code}}

Since you proposed the original fix, explain how this transformed buggy function should be fixed. 
\end{lstlisting}


\begin{lstlisting}[language=LLMPrompt, caption={Follow-up Explanation Verification}]
System: You are an Automatic Program Repair Tool. You are now the examinee in a debate-style viva examination, where you defend your original repair decision.
Answer YES or NO. If you are NOT SURE, answer NOT SURE. Don't return anything else except YES, NO, or NOT SURE.
User:
// Buggy Function
{{original_buggy_code}}

// Explanation
{{explanation}}

Given a buggy function and an explanation, you need to verify if the explanation is applicable to fix the buggy function.
\end{lstlisting}

\section{Annotation Process}
\label{sec:AppendixManualLabel}

Experts were invited to label the correctness of 1,756 explanations, as collected in Section~\ref{sec:benchmark}.
Each has 3--5 years of experience in programming and program repair.
They first went through a tutorial on how to review and label an explanation.
\begin{figure*}
    \centering
    \includegraphics[width=\linewidth, clip, trim=0 0 0 1cm]{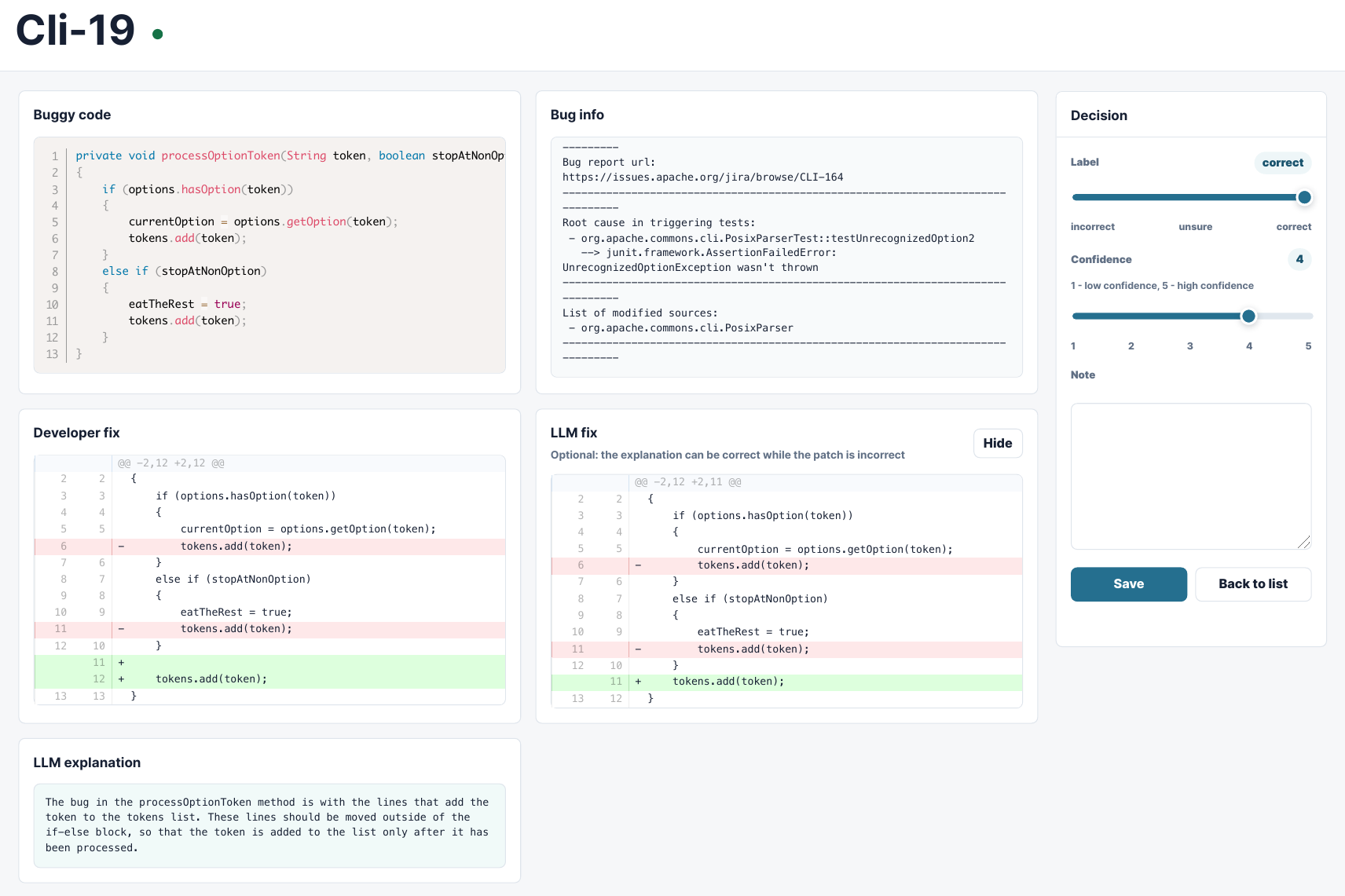}
    \caption{Annotation page of an MORepair-generated explanation for Cli-19 in Defects4J.}
    \label{fig:AppendixManualLabel}
\end{figure*}
Then, for each explanation, two annotators were assigned to independently label its correctness.
A web application was developed to facilitate this process.
Figure~\ref{fig:AppendixManualLabel} shows a screenshot of the web application for labeling an explanation generated by MORepair for Cli-19 in Defects4J.
The annotators were provided with various artifacts, including the buggy program, the developer's fix, the LLM-generated explanation, the LLM-generated fix, and auxiliary information about the bug, such as the bug report and triggering tests.
After reviewing these artifacts, the annotators decided whether the explanation is correct and stated their level of confidence from 1 to 5 on a Likert-like scale. 
They were also encouraged to note their rationale during the labeling process.
If they are unsure, the third option ``unsure'' could be selected.

For example, the program \textcode{processOptionToken} in Figure~\ref{fig:AppendixManualLabel} does not handle the token properly.
The token is simply ignored if it does not belong to \textcode{options} or if \textcode{stopAtNonOption} is disabled.
Based on the developer's fix, this buggy program can be fixed by moving \textcode{tokens.add(token)} outside of the if-else block.
The explanation generated by MORepair correctly describes the bug and how to fix it.
As a result, this explanation is labeled as correct.

The web application recorded that it took 182.62 seconds on average for an annotator to label an explanation.
In total, it took 178.15 hours to label the entire 1,756 explanations.
This annotation process achieved a Cohen's Kappa value for inter-annotator agreement of 0.78.
This indicates substantial consistency and reliability, significantly bolstering our confidence in both the benchmark's quality and \MT's evaluation.

If an annotator is unsure about the explanation's correctness or the two annotators disagree, a third annotator got involved in the discussion to reach a consensus.
Ultimately, 1,217 incorrect explanations were identified, as shown in Table~\ref{table:overall_benchmark}.
This shows that many generated explanations, in particular 69.3\% of the explanations, are incorrect. 
The APR systems are still highly unreliable for explaining their repairs, even the best-performing system, ReInFix.

\begin{table*}[t]
\centering
\caption{Benchmarking the correctness of LLM-generated APR explanations}
\label{table:breakdownbench}
\begin{tabular}{l|r|r|r|r|p{3cm}|r|r|r|r}
\toprule
& \multicolumn{4}{c|}{Defects4J} & \multicolumn{5}{c}{Other Datasets} \\
\textbf{APR System} & & \textbf{\# Correct} & \textbf{\# Incorrect} & \textbf{Incorrect} &  & & \textbf{\# Correct} & \textbf{\# Incorrect} & \textbf{Incorrect} \\
\textbf{Under Test} & \textbf{\# Expls} & \textbf{Expls} & \textbf{Expls} & \textbf{Expl Rate} & \textbf{Name} & \textbf{\# Expls} & \textbf{Expls} & \textbf{Expls} & \textbf{Expl Rate} \\
\midrule
ThinkRepair & 369 & 44 & 325 & 88.1\% & RWB & 62 & 8 & 54 & 87.1\% \\
\midrule
ReInFix & 670 & 287 & 383 & 57.2\% & RWB & 41 & 18 & 23 & 56.1\% \\
\midrule
MORepair & 281 & 58 & 223 & 79.4\% & SWE-Repair, EvalRepair-C++, EvalRepair-Java & 333 & 122 & 211 & 63.4\% \\
\bottomrule
\end{tabular}
\end{table*}

\section{Experiments}

\subsubsection{Sampling Transformed Programs} \label{sec:sample_programs}

The generations of transformed programs $\text{\textcode{Tran}}_{\text{\textit{eqv}}}$ and $\text{\textcode{Tran}}_{\text{\textit{neq}}}$ are typically inexpensive.
However, generating follow-up explanations $\text{\textcode{Explain}}_m$ may be costly, as it requires querying the LLM multiple times. 
To reduce this cost, we adopt a fixed sampling budget strategy: given a budget $N$, we draw $N$ follow-up explanations for $N$ equivalent programs and $N$ follow-up explanations for $N$ non-equivalent ones.
To enhance the diversity of this sampling, \MT{} selects the most representative transformed programs.

To do this, \MT{} clusters each list of equivalent and non-equivalent transformed programs into $N$ clusters.
\MT{} transform each program into its vector representation using UniXCoder\footnote{https://github.com/microsoft/CodeBERT/tree/master/UniXcoder}.
For each category of equivalence or non-equivalence, \MT{} clusters the vector representations of transformed programs into $N$ clusters.
Then, it randomly chooses one in each cluster as a representative example.

\begin{table}[ht]
\caption{Comparing two strategies for sampling transformed programs: clustering and random selection}
\label{table:sample_transformed_programs}
\setlength{\tabcolsep}{5pt} 
\begin{tabular}{c|l|r|r|r|r|r|r}
\toprule
\textbf{System} & \multirow{2}{*}{\textbf{Detector}} & \multicolumn{6}{c}{\textbf{Detection Quality}} \\
\textbf{Under Test} & & \textbf{MCC} & \textbf{$\Delta$MCC} & \textbf{BA} & \textbf{Pre} & \textbf{Rec} & \textbf{Spe} \\
\midrule
\multirow{2}{*}{ThinkRepair} & Cluster & 0.34 &  & 0.75 & 0.96 & 0.72 & 0.77 \\
 & Random & 0.04 & \textcolor{red}{-88.2\%} & 0.52 & 0.88 & 0.84 & 0.20 \\
\midrule
\multirow{2}{*}{ReInFix} & Cluster & 0.43 &  & 0.70 & 0.88 & 0.50 & 0.91 \\
 & Random & 0.26 & \textcolor{red}{-39.5\%} & 0.61 & 0.79 & 0.36 & 0.87 \\
\midrule
\multirow{2}{*}{MORepair} & Cluster & 0.39 &  & 0.71 & 0.91 & 0.70 & 0.72 \\
 & Random & 0.15 & \textcolor{red}{-61.5\%} & 0.58 & 0.76 & 0.63 & 0.53 \\
\bottomrule
\end{tabular}
\end{table}
Table~\ref{table:sample_transformed_programs} compares \MT's effectiveness with two strategies for sampling transformed programs.
The default strategy is clustering, as aforementioned.
The alternative strategy is to randomly select $N$ programs.
The results show that clustering is consistently more effective than random selection for sampling transformed programs across all three APR systems.
For ThinkRepair, MCC drops from 0.42 with clustering to only 0.04 with random selection, a reduction of 91.4\%. 
ReInFix also drops from 0.47 to 0.24, while MORepair decreases from 0.39 to 0.14. 
This shows that random sampling substantially weakens \MT's ability to distinguish correct and incorrect explanations.
Precision, Recall, and Specificity help clarify the difference between the two strategies. 
For ThinkRepair, random selection achieves higher Recall than clustering, but its Specificity drops sharply from 0.75 to 0.20. 
This means \MT{} correctly detects many incorrect explanations but incorrectly labels correct ones as incorrect, which severely reduces MCC. 
For ReInFix, the opposite trend appears. 
Random selection has slightly higher Specificity, but Recall falls from 0.67 to 0.36, meaning it misses many true incorrect explanations. 
For MORepair, random selection performs worse on almost every metric.
Overall, the results suggest that clustering selects a more diverse and representative set of transformed programs, leading to more effective detection. 

\subsubsection{Interpretation of Evaluation Metrics}
Tables~\ref{table:InterpretMCC}--\ref{table:InterpretKappa} show the interpretations of Matthew Correlation Coefficient (MCC)~\cite{Schober:2018:CorrelationCoefficients:AppropriateUseandInterpretation} and Cohen's Kappa~\cite{TheKappaStatistic}, respectively.

\begin{table}[ht]
\centering
\caption{Interpretation of Matthew Correlation Coefficient (MCC), based on thresholds from~\cite{Schober:2018:CorrelationCoefficients:AppropriateUseandInterpretation}}
\label{table:InterpretMCC}
\begin{tabular}{p{2.6cm}|l}
\toprule
\textbf{Matthew Correlation Coefficient (MCC)} & \textbf{Interpretation} \\
\midrule
0.00 -- 0.09 & Negligible correlation \\
0.10 -- 0.39 & Weak correlation \\
0.40 -- 0.69 & Moderate correlation \\
0.70 -- 0.89 & Strong correlation \\
0.90 -- 1.00 & Very strong correlation \\
\bottomrule
\end{tabular}
\vspace{5mm}
\caption{Interpretation of Cohen's Kappa~\cite{TheKappaStatistic}}
\label{table:InterpretKappa}
\centering
\begin{tabular}{l|l}
\toprule
\textbf{Value of Kappa} & \textbf{Level of Agreement} \\
\midrule
0.00 -- 0.20 & None \\
0.21 -- 0.39 & Minimal \\
0.40 -- 0.59 & Weak \\
0.60 -- 0.79 & Moderate \\
0.80 -- 0.90 & Strong \\
0.91 -- 1.00 & Almost Perfect \\
\bottomrule
\end{tabular}
\end{table}

\subsubsection{Benchmark Breakdown}
\label{sec:AppendixResults}

Table~\ref{table:breakdownbench} shows that incorrect explanations are common across all APR systems and datasets. 
On Defects4J, all three systems produce a large proportion of incorrect explanations.
ThinkRepair has the highest incorrect rate at 88.1\%, followed by MORepair at 79.4\%, and ReInFix at 57.2\%. 
Even the best-performing system, ReInFix, still produces more incorrect explanations than correct ones.

The trend remains consistent on the other datasets. 
ThinkRepair and ReInFix show almost the same incorrect rates as on Defects4J, with 87.1\% and 56.1\%, respectively. 
This suggests that their explanation correctness is relatively stable across bug benchmarks. 
MORepair achieves a lower incorrect rate on the other datasets, dropping from 79.4\% to 63.4\%, but incorrect explanations still dominate.

Overall, these results suggest that incorrect APR explanations are not an isolated issue of Defects4J. 
Across all datasets, there are 1,756 explanations, of which 1,219 are incorrect, giving an overall incorrect explanation rate of about 69.3\%. 
This supports the need for automated methods such as MetaReJ to detect incorrect explanations before they are presented to developers.

\subsubsection{Detection Examples}
\label{sec:AppendixExamples}

We provide an example of a true positive, where an incorrect explanation is correctly detected. 
\begin{lstlisting}[language=java,label=lst:Lang40,caption={The buggy program \textcode{containsIgnoreCase}, also known as Lang-40 in Defects4J.}]
public static boolean containsIgnoreCase(String str, String searchStr) {
    if (str == null || searchStr == null) {
        return false;
    }
    return contains(str.toUpperCase(), searchStr.toUpperCase());
}
\end{lstlisting}

\begin{lstlisting}[language=java,caption={The fix for Lang-40 generated by ThinkRepair.}, numbers=none]
-   return contains(str.toUpperCase(), searchStr.toUpperCase());
+   return contains(str.toUpperCase(), searchStr.toUpperCase()) || contains(str.toLowerCase(), searchStr.toLowerCase());
\end{lstlisting}
Original Explanation: The bug in the containsIgnoreCase function is with the line `contains(str.toUpperCase(), searchStr.toUpperCase())`. This line checks if the search string is a substring of the input string, but it is incorrect and will result in a stack overflow for certain inputs.

\begin{lstlisting}[language=java,caption={A transformed program semantically non-equivalent to Lang-40 in Defects4J. The logical operator \textcode{||} is replaced by \textcode{\&\&}.}]
public static boolean containsIgnoreCase(String str, String searchStr) {
    if (null == str && null == searchStr) {
        return false;
    }
    return contains(str.toUpperCase(), searchStr.toUpperCase());
}
\end{lstlisting}

Follow-up Explanation: The bug in the original function is that it returns `false` if either `str` or `searchStr` is `null`, even if the other argument is not `null`. This is because the `\&\&` operator short-circuits and returns `false` as soon as it encounters a `null` value.
To fix this bug, we need to change the `\&\&` operator to `$||$` so that both arguments are checked for `null` before calling the `contains` method. This will ensure that the function returns `false` only if both `str` and `searchStr` are `null`.

This follow-up explanation explains that we need to change the `\&\&` operator to `$||$`, which is not applicable to the original Lang-40.
However, ThinkRepair responds that it is not applicable.

We provide an example of a false negative.
\begin{lstlisting}[language=java,label=lst:Lang12,caption={The buggy program \textcode{random}, also known as Lang-12 in Defects4J.}]
public static String random(int count, int start, int end, boolean letters, boolean numbers, char[] chars, Random random) {
  if (count == 0) {
    return "";
  } else if (count < 0) {
    throw new IllegalArgumentException("Requested random string length " + count + " is less than 0.");
  }

  if (start == 0 && end == 0) {
      if (!letters && !numbers) {
        end = Integer.MAX_VALUE;
      } else {
        end = 'z' + 1;
        start = ' ';
      }
  }

  char[] buffer = new char[count];
  int gap = end - start;

  while (count-- != 0) {
    char ch;
    if (chars == null) {
      ch = (char) (random.nextInt(gap) + start);
    } else {
      ch = chars[random.nextInt(gap) + start];
    }
    if (letters && Character.isLetter(ch)
        || numbers && Character.isDigit(ch)
        || !letters && !numbers) {
      if(ch >= 56320 && ch <= 57343) {
        if(count == 0) {
          count++;
        } else {
          // low surrogate, insert high surrogate after putting it in
          buffer[count] = ch;
          count--;
          buffer[count] = (char) (55296 + random.nextInt(128));
        }
      } else if(ch >= 55296 && ch <= 56191) {
        if(count == 0) {
          count++;
        } else {
          // high surrogate, insert low surrogate before putting it in
          buffer[count] = (char) (56320 + random.nextInt(128));
          count--;
          buffer[count] = ch;
        }
      } else if(ch >= 56192 && ch <= 56319) {
        // private high surrogate, no effing clue, so skip it
        count++;
      } else {
        buffer[count] = ch;
      }
    } else {
      count++;
    }
  }
  return new String(buffer);
}
\end{lstlisting}

\begin{lstlisting}[language=java,caption={The fix for Lang-12 generated by ThinkRepair.},numbers=none]
-   if (start == 0 && end == 0) {
+   if (start == 0 || end == 0) {
\end{lstlisting}
Original Explanation: The bug is in the `if (start == 0 \&\& end == 0)` statement. The function is checking if both `start` and `end` are 0, but the intended behavior is to check if either one of them is 0.

This repair generated by ThinkRepair is incorrect.
The actual bug is that the chars array may be null or empty and there is currently no check for that.
The repair incorrectly identifies the root cause, and proposes a wrong explanation.

We transform the program Lang-12 into a semantically equivalent program as follows.
\begin{lstlisting}[language=java,caption={A transformed program semantically equivalent to Lang-12 in Defects4J.}]
public static String random(int count, int start, int end, boolean letters, boolean numbers, char[] chars, Random random) {
  if (count < 0) {
    throw new IllegalArgumentException("Requested random string length " + count + " is less than 0.");
  } else if (count == 0) { 
    return "";
  }

  if (start == 0 && end == 0) {
      if (!letters && !numbers) {
        end = Integer.MAX_VALUE;
      } else {
        end = 'z' + 1;
        start = ' ';
      }
  }

  char[] buffer = new char[count];
  int gap = end - start;

  while (count-- != 0) {
    char ch = (chars == null) ? (char) (random.nextInt(gap) + start) : chars[random.nextInt(gap) + start];
    if (letters && Character.isLetter(ch)
        || numbers && Character.isDigit(ch)
        || !letters && !numbers) {
      if(ch >= 56320 && ch <= 57343) {
        if(count == 0) {
          count++;
        } else {
          // low surrogate, insert high surrogate after putting it in
          buffer[count] = ch;
          count--;
          buffer[count] = (char) (55296 + random.nextInt(128));
        }
      } else if(ch >= 55296 && ch <= 56191) {
        if(count == 0) {
          count++;
        } else {
          // high surrogate, insert low surrogate before putting it in
          buffer[count] = (char) (56320 + random.nextInt(128));
          count--;
          buffer[count] = ch;
        }
      } else if(ch >= 56192 && ch <= 56319) {
        // private high surrogate, no effing clue, so skip it
        count++;
      } else {
        buffer[count] = ch;
      }
    } else {
      count++;
    }
  }
  return new String(buffer);
}
\end{lstlisting}
The transformed program makes two refactoring changes while preserving the original behavior. 
First, it reorders the initial boundary checks by checking `count < 0` before `count == 0`; this does not change the logic because the two conditions are mutually exclusive. Second, it replaces the explicit `if-else` block used to assign `ch` with an equivalent ternary expression. 
Thus, the transformation simplifies the code structure and makes the assignment more compact, but it does not alter the program’s intended functionality.

Follow-up Explanations: The bug is caused by the fact that the function is using the `\&\&` operator instead of the `$||$` operator to check if either `start` or `end` is equal to 0. This means that the function will only set `end` to `Integer.MAX\_VALUE` if both `start` and `end` are equal to 0, which is not the intended behavior. To fix this bug, you should change the line `if (start == 0 \&\& end == 0) \{ ... \}` to `if (start == 0 $||$ end == 0) \{ ... \}`. This will ensure that the function sets `end` to `Integer.MAX\_VALUE` if either `start` or `end` is equal to 0, which is the intended behavior.

This follow-up explanation is similar to the original one.
Although the follow-up explanation is incorrect, as the original, the LLM verifies that it is applicable to the buggy program Lang-12.

We also transform the program Lang-12 into a semantically non-equivalent program as follows.
\begin{lstlisting}[language=java,caption={A transformed program semantically non-equivalent to Lang-12 in Defects4J.}]
public static String random(int count, int start, int end, boolean letters, boolean numbers, char[] chars, Random random) {
  char[] buffer = new char[count];
  int gap = end - start;

  while (count-- != 0) {
    char ch;
    if (chars == null) {
      ch = (char) (random.nextInt(gap) + start);
    } else {
      ch = chars[random.nextInt(gap) + start];
    }
    if (letters && Character.isLetter(ch)
        || numbers && Character.isDigit(ch)
        || !letters && !numbers) {
      if(ch >= 56320 && ch <= 57343) {
        if(count == 0) {
          count++;
        } else {
          // low surrogate, insert high surrogate after putting it in
          buffer[count] = ch;
          count--;
          buffer[count] = (char) (55296 + random.nextInt(128));
        }
      } else if(ch >= 55296 && ch <= 56191) {
        if(count == 0) {
          count++;
        } else {
          // high surrogate, insert low surrogate before putting it in
          buffer[count] = (char) (56320 + random.nextInt(128));
          count--;
          buffer[count] = ch;
        }
      } else if(ch >= 56192 && ch <= 56319) {
        // private high surrogate, no effing clue, so skip it
        count++;
      } else {
        buffer[count] = ch;
      }
    } else {
      count++;
    }
  }
  return new String(buffer);
}
\end{lstlisting}
This new program removes the if statement `if (start == 0 \&\& end == 0) \{ ... \}`.
ThinkRepair provides a follow-up explanation for this program.

Follow-up Explanation: The bug in the original function is that it does not correctly handle the case where `start` and `end` are both 0. In this case, the function sets `end` to `Integer.MAX\_VALUE` and `start` to `' '`, which is not the correct behavior. 
To fix this bug, we need to check if `start` and `end` are both 0, and if so, set `end` to `Integer.MAX\_VALUE` and `start` to `0`. We can do this by adding the following code at the beginning of the function:
\begin{lstlisting}[language=java, numbers=none]
if (start == 0 && end == 0) {
    end = Integer.MAX_VALUE;
    start = 0;
}
\end{lstlisting}
This follow-up explanation is completely different from the original. 
ThinkRepair verifies that this explanation is not applicable to the original program.
In this case, ThinkRepair is stubborn and consistently defends its wrong reasoning.
As a result, \MT{} fails to detect the original explanation generated by ThinkRepair as incorrect.